\documentclass{aa}
\usepackage{graphicx}
\usepackage{txfonts}
\begin{document}
   \title{Stellar population analysis on local infrared-selected galaxies}

   \author{X. Y. Chen \inst{1,2}\thanks{email: chenxy@bao.ac.cn},
          Y. C. Liang\inst{1}\thanks{email: ycliang@bao.ac.cn}, F. Hammer\inst{3}, Y. H. Zhao\inst{1}, G. H. Zhong\inst{1,4}
          }
   \institute{
National Astronomical Observatories,
                Chinese Academy of Sciences, A20 Datun Road, 100012 Beijing, China
         \and
 Graduate School of the Chinese Academy of Sciences, 100049 Beijing, China
         \and
 GEPI, Observatoire de Paris-Meudon, 92195 Meudon, France
         \and
Department of Physics, Hebei Normal University, 050016 Shijiazhuang, China
             }

   \date{Received ; accepted }

% \abstract{}{}{}{}{}
% 5{} token are mandatory

  \abstract
  %{} leave it empty if necessary
{}
{We study the stellar population of local infrared galaxies, which
contain star-forming galaxies, composite galaxies, LINERs, and
Seyfert 2s. We also want to  find whether infrared luminosity and
spectral class have any effect on their stellar populations.
   }
{The sample galaxies are selected from the main galaxy sample of
SDSS-DR4 and then cross-correlated with the IRAS-PSCz catalog. We
fit our spectra (stellar absorption lines and continua) using the
spectral synthesis code STARLIGHT based on the templates of simple
stellar population and the spectra of star clusters.
   }
{Among the 4 spectral classes, LINERs present the oldest stellar
populations, and the other 3 subsamples all present substantial
young and intermediate age populations and very few old populations.
The importance of young populations decreases from star-forming,
composite, Seyfert 2 to LINER. As for the different infrared
luminosity bins, ULIGs \& LIGs (log($L_{IR}/L_{\odot})\geq$11)
present younger populations than do starbursts and normal galaxies;
however, the dominant contributors to mass are old populations in
all sample galaxies. The fittings also give consistent results by
using the spectra of star clusters with different ages and
metallicities as templates. The dominated populations in
star-forming and composite galaxies are those with metallicity
$Z=0.2Z_\odot$, while LINERs and Seyfert 2s are more metal-rich. The
normal galaxies are more metal-rich than the ULIGs \& LIGs and
starbursts for the star-forming galaxies within different infrared
luminosity bins.

Additionally, we compare some synthesis results with other
parameters obtained from the MPA/JHU catalog. First we find that the
stellar and nebular extinctions are correlated, and the ionized gas
suffers twice as much extinction as stars. Second we confirm that
$D_{n}(4000)$ is a much better age indicator than $H\delta_{A}$.
Following the evolution of galaxies, $D_{n}(4000)$ monotonously
varies.
 Finally we investigate some relationships between mean
stellar age, mean stellar metallicity, and nebular metallicity for
the subsample of star-forming galaxies. In star-forming galaxies,
the nebular metallicity $Z_{neb}$ is correlated with the
light-weighted mean stellar age $\langle\,log\,t_{\ast}\rangle_{L}$
in an intermediate strength, and $Z_{neb}$ is weakly correlated with
the mass-weighted mean stellar metallicity
$\langle\,Z_{\ast}\rangle_{M}$.
  }
{}
\authorrunning{X. Y. Chen et al.}
\titlerunning{Stellar population analysis on local infrared-selected galaxies}

\maketitle

%==================================================================
\section{Introduction}
\label{sec.1} Infrared galaxies are objects that emit most of their
energy in the infrared ($\sim 5-500\mu\,m)$ (Sanders \& Mirabel
1996). They are one of the most important discoveries from
extragalactic observations at mid- and far-infrared wavelengths. The
overwhelming majority of the extragalactic objects detected in the
Infrared Astronomical Satellite (IRAS) survey are late-type spiral
galaxies, while elliptical and S0 galaxies are rarely detected
(Soifer 1987). With the higher sensitivity and spatial resolution
than IRAS, the ISO satellite moves the studies of luminous infrared
galaxies to $z\geq$0.5 from local (Elbaz et al. 1999, 2002; Flores
et al. 1999; Liang et al. 2004; Zheng et al. 2004; Hammer et al.
2005 and references therein). The cosmic infrared background
resolved by ISO mid-infrared camera (ISOCAM) shows that the
co-moving density of infrared light due to the luminous IR galaxies
was more than 40-70 times higher at $z\sim 1$ than today (Elbaz et
al. 2002). And this evolution is mainly driven by the luminous
infrared starburst galaxies, in which the star formation rate runs
up to $100M_{\odot}\,yr^{-1}$ (Flores et al. 1999). The successful
launch of the Spitzer Space Telescope opened a new exciting window
on the deep infrared universe. Spitzer operates between 3.6 and 160
$\mu$m with better sensitivity and spatial resolution than for
IRAS and ISO, and provides new opportunities for determining the IR
properties of galaxies in the general context of cosmic evolution (
Le Floc'h et al. 2005, Zheng et al. 2007; Papovich et al. 2007 and
references therein). The future Herschel will also improve much of
our knowledge about infrared galaxies. Therefore, infrared-detected
galaxies are one of the most interesting objects in the Universe and
are related to the major star-forming process. The local infrared
galaxies have the advantage of being studied in details with
properties, such as stellar populations, since their high-quality
optical spectra could be obtained in the modern digital sky survey,
e.g. the Sloan Digital Sky Survey (SDSS).

Understanding the overall stellar population in galaxies is a
crucial tool for unveiling the star formation and evolution of
galaxies. As spectra of galaxies record information about age and
metallicity distributions of its stellar populations, many works
have been generated on stellar population analysis through spectra
of galaxies, using the approach called stellar population synthesis.
The first attempts were the empirical population synthesis (Faber
1972), in which the observed spectrum of a galaxy is reproduced by a
combination of spectra of individual stars or star clusters of
different ages and metallicities from a library. More recent models
have been based on the evolutionary population synthesis technique
(Tinsley 1978; Bruzual 1983; Charlot et al. 1996; Bruzual \& Charlot
2003), in which the main adjustable parameters could be the stellar
initial mass function, the star formation rate, and the chemical
enrichment. The reproduced simple stellar populations (SSP) with
different ages and metallicities from the evolutionary population
synthesis model could be used to analyze the stellar populations of
the galaxies.

Some research has taken on analysis of stellar populations in
star-forming galaxies and starburst galaxies from the optical band
observations, most of which conclude that H\,II galaxies are
age-composite stellar systems (Schmitt et al. 1996; Raimann et al.
2000; Cid Fernandes et al. 2003; Kong et al. 2003; Westera et al.
2004). Hammer et al. (2001) used the spectra of star clusters to
analyze the stellar populations of distant luminous blue compact
galaxies at redshift $0.4<z<1$ and find that they have included some
old populations.

Moreover, some works have been performed on ultraviolet (UV) and near
infrared bands. Based on International Ultraviolet Explore (IUE) UV
spectra, Bonatto et al. (2000) discuss the stellar population of
galaxies with enhanced star formation, which are at distances
corresponding to radial velocities in the range of 5,000-16,000 km
s$^{-1}$. They find that young stellar populations (age $<$ 500 Myr)
are the main contributors except for galaxies with a red spectrum
arising from the intermediate (age $\sim$ 1-2 Gyr) and old-age
populations. Riffel et al. (2008) have studied the central stellar
populations of four starburst galaxies through near-infrared
spectroscopy and find that the dominate population was a 1Gyr old
with solar metallicity component in all galaxies.

Furthermore, some research was applied to AGNs. Cid Fernandes et al.
present a series of studies that discussed the stellar populations
of low-luminosity active galactic nuclei (LLAGN) (Cid Fernandes et
al. 2004a, 2005a; Gonz\'{a}lez  Delgado et al. 2004). They find that
young stars contribute very little to the optical continuum, while
intermediate-age stars contribute significantly; and most of the
strong-[OI] LLAGNs predominantly have old stellar population.
Besides that, some works have been dedicated to the stellar
populations of Seyfert 2s, and they confirm that most Seyfert 2s
contain a young hot stellar population (Schmitt et al. 1999;
Storchi-Bergmann et al. 2000; Gonz\'{a}lez Delgado et al. 2001;
Joguet et al. 2001; Cid Fernandes et al. 2004b). Boisson et al.
(2000) performed stellar population synthesis on a sample of 12
galaxies (Boisson et al. 2004 studied 5 more) with different levels
of activity, and they find different populations for different
activity types: the starburst galaxies present the youngest
populations of the sample; the Seyfert 2s also have young stellar
population, less intense than in the starbursts, but metal rich;
while LINERs show the oldest populations, metal rich, with little
star formation still going on. But their sample was small, so a much
larger sample from the modern digital sky survey is needed to verify
these results.

To our knowledge, no work has been specifically dedicated to the
stellar population analysis of local infrared-selected galaxies. In
this paper, we utilize the data from cross identifying between SDSS
and IRAS to study the stellar population of local (with median value
of redshift $\sim 0.03$) infrared-selected galaxies by using a
stellar population synthesis program called STARLIGHT (Cid Fernandes
et al. 2005b). With the high-quality SDSS optical spectra and the
IRAS observations, this could be the largest local sample selected
from infrared to date for studying their stellar populations from
the full spectrum fitting pixel by pixel. Moreover, as our sample
contains diverse spectral classes (star-forming galaxies, composite
galaxies, LINERs, and Seyfert 2s) and spans a broad range of
infrared luminosity, we can compare the properties among these
different classes.

 This paper is organized as follows.
We present the sample selection and the classifications in Sect.\,2.
In Sect.\,3 we describe the methods of the spectral synthesis and
the results of fitting the spectra of galaxies in different classes
and different bins of infrared luminosity. In Sect.\,4, we present
the relations of some property parameters from emission lines,
continua of galaxies, and some from the spectral synthesis fittings
on the spectra. Summary and conclusions are given in Sect.\,5.
Throughout this paper we assume the following cosmological
parameters: $\Omega_{M}=0.3$, $\Omega_{\Lambda}=0.7$, and $H_{0}$=70
km\,s$^{-1}$\,Mpc$^{-1}$.

%==============================================================
\section{Sample selection and classification}
\label{sec.2} Our sample was selected from the main galaxy sample
of the Sloan Digital Sky Survey (SDSS) DR4 database and
then was cross-correlated with the
IRAS Point Source
Catalog (PSCz).

The SDSS-DR4 includes over 800,000 fiber spectra of objects selected
based on the five bands ($u,g,r,i$, and $z$) imaging (Fukugita et
al. 1996; Smith et al. 2002; Strauss et al. 2002). Each fiber is
$3\arcsec$  in diameter (Bernardi et al. 2003; Greene et al. 2005,
2006; Adelman-McCarthy et al. 2006), corresponding to $\sim 1.8$ kpc
at $z=0.03$ (the median redshift of our sample, see below and
Fig.~\ref{849zlirdis.eps}). The observed-frame spectral wavelength
range is 3800-9200 \AA. The instrumental resolution of the spectra
is $\lambda/\Delta \lambda \approx 1800$ equivalent to Gaussian
$\sigma_{\rm inst} \approx 70\,{\rm km\,s^{-1}}$ (Heckman et al.
2004; Greene et al. 2005, 2006).

In particular, we selected all galaxies with redshift less than
$0.5$ and redshift confidence greater than $0.9$ from SDSS DR4.
After that these objects were cross-identified with PSCz using a
$5\arcsec$ matching radius. The IRAS PSCz survey is a redshift
survey of about 15,000 galaxies detected, after supplements and
corrections (Saunders et al. 2000), containing 18,351 objects with
$f_{60}>0.6 Jy$, covering 84\% of the full sky. The optical
positions of PSCz objects were used to assure the reliability of the
identification. As the PSCz objects mostly have redshift less than
$0.1$ (Saunders et al. 2000), the redshift distribution of our
sample objects is almost truncated after a redshift of $0.1$.

After cross-identification, 1097 target objects were obtained. To
improve the accuracy of the classification of our sample objects, we
excluded 34 objects without emission-line flux measured in the
MPA/JHU catalog. Meanwhile we excluded another 65 objects with low
emission-line signal-to-noise ratio (S/N) according to their ${\rm
H}{\beta},{\rm H}{\alpha}$, and [NII]$\lambda6583$ lines detected at
lower than 5$\sigma$, and [OIII]$\lambda5007$ line detected at
lower than 3$\sigma$. After this cut, 998 sources are left.

To ensure the spectra were taken at the centers of the galaxies (within a radius
of $2\arcsec$ from the center),
we examined the positions of fibers for all the sample objects.
We found that there were 49 galaxies
with the fibers not located at the centers. After removing them,
we have 949 objects left in total.

 To do the spectral synthesis fittings, 100 objects were further
ejected because their mask spectra were problematic, in which almost
half of the spectra in wavelength were masked (see more details in
Sect.~\ref{sec.3.1}). Finally, 849 objects were left as our working
sample for a stellar population analysis.

 The redshifts of the whole sample of 849 galaxies range from
0.002 to 0.24, with a median of 0.03, which is presented in
Fig.~\ref{849zlirdis.eps} (left column). In this work, we used two
methods (by emission-line ratios and by infrared luminosity) to
classify the sample galaxies into several subgroups as shown below.

%==========================================================
\subsection{Classification by emission-line ratios}
\label{sec.2.1} The first method to classify the samples is using an
emission-line diagnostic diagram. Namely, we classified the 849
sources as star-forming galaxies, composite galaxies, LINERs, and
Seyfert 2s according to the emission-line flux ratio diagnostic
diagram (Fig.~\ref{849bpt.eps}, Baldwin et al. 1981, BPT; Veilleux
\& Osterbrock 1987; Kauffmann et al. 2003a; Kewley et al. 2001).
First, we separated star-forming galaxies, composite galaxies, and
AGNs by Eq.~(\ref{k01}) (Kewley et al. 2001, Fig.~\ref{849bpt.eps})
and Eq.~(\ref{k03}) (Kauffmann et al. 2003a, Fig.~\ref{849bpt.eps}).
The composite galaxies refer to objects whose spectra contain
significant contributions from both AGN and star formation
(Brinchmann et al. 2004, Kewley et al. 2006).
 Then we
divided AGNs into LINERs and Seyfert 2s by Eq.~(\ref{s81}) (Shuder
et al. 1981). The fluxes of these emission lines were derived from
the MPA/JHU catalog. We finally obtained 419 star-forming galaxies,
326 composite galaxies, 69 LINERs, and 35 Seyfert 2s
(Table~\ref{number}):

\begin{equation} \label{k01}
\log\left (\frac{\mathrm{[OIII]}\lambda5007}{H\beta}\right)
=\frac{0.61}{\log (\mathrm{[NII]\lambda6583/H\alpha}) -0.47}+1.19,
\end{equation}

\begin{equation} \label{k03}
\log\left
(\frac{\mathrm{[OIII]}\lambda5007}{\mathrm{H\beta}}\right)
=\frac{0.61}{\log (\mathrm{[NII]\lambda6583/H\alpha}) -0.05}+1.3,
\end{equation}

\begin{equation}
\label{s81} \rm{[OIII]}\lambda5007/H\beta=3.
\end{equation}

%==============================================================
\subsection{Classification by infrared luminosity}
\label{sec.2.2}
\begin{figure}
   \centering
   \includegraphics[width=8cm]{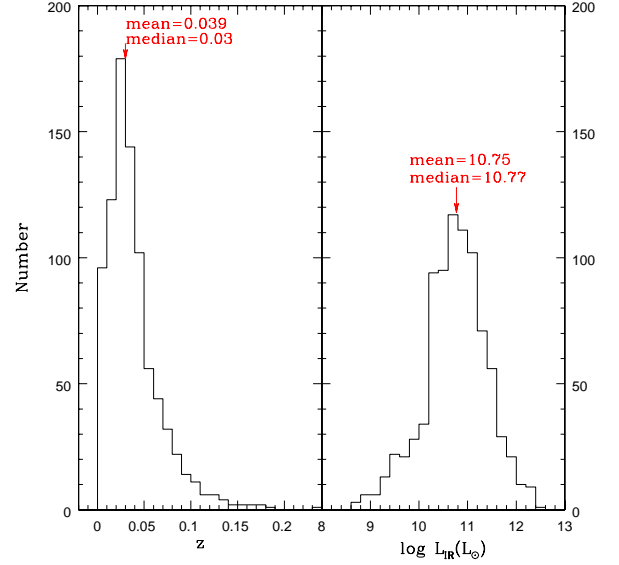}
   \caption{The distributions of redshift (left) and infrared luminosity (right)
for our sample of 849 infrared-selected galaxies.
           }
   \label{849zlirdis.eps}
    \end{figure}

\begin{figure}
\centering
   \includegraphics[width=8cm]{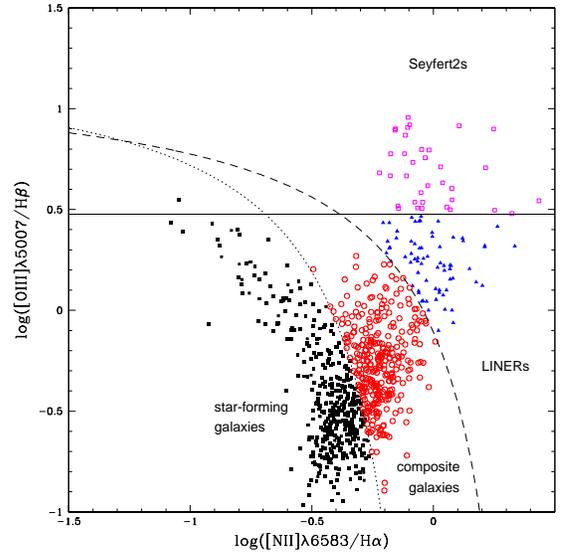}
   \caption{The emission-line flux ratio diagnostic diagram (i.e. BPT, Baldwin et al. 1981)
for our full sample (849), in which we plot the ratio of
$[OIII]/H\beta$ versus the ratio of $[NII]/H\alpha$ . The dotted
curve defined by Kauffmann et al. (2003a) and the dashed curve
defined by Kewley et al. (2001) show the separation among
star-forming galaxies, composite galaxies, and AGNs. The solid
line defined by Shuder et al. (1981) shows the separation between
LINERs and Seyfert 2s.}
   \label{849bpt.eps}
    \end{figure}

We calculated the far-infrared luminosity $L_{\rm FIR}$ of the sample galaxies
from the nominal flux densities, $f_{\nu}$(60 $\mu${\rm m}) and
$f_{\nu}$(100 $\mu${\rm m}) in the IRAS catalogs (Helou et al.
1988; Sanders \& Mirabel 1996), and the redshifts given by SDSS,
and then we converted it to the total infrared luminosity $L_{\rm
IR}$ (1-1000 $\mu$m, Calzetti et al. 2000; Cao et al. 2006; Wang
et al. 2006; Wang 2008; Chen et al. 2008) according to the following formulas:

\begin{equation}
\label{lir1} F_{\rm FIR}=1.26 \times 10^{-14}
\{2.58f_{60}+f_{100}\}{\rm [W\,m^{-2}]},
\end{equation}
\begin{equation}
\label{lir2} L_{\rm FIR}=4 \pi D_{\rm L}^{2} F_{\rm FIR}
[L_{\odot}],
\end{equation}
\begin{equation}
\label{lir3} L_{\rm IR}(1\sim1000 \mu_{\rm m})=1.75L_{\rm FIR}.
\end{equation}
The infrared luminosity of the whole sample of 849 galaxies spans a
range of $\log (L_{\rm IR}/L_{\odot}) =8.65-12.53$, with a median of
10.77, and we show the distribution of infrared luminosity in
Fig.~\ref{849zlirdis.eps} (right column).

The second classification method was in accordance with their
infrared luminosities. Namely, we divided our 849 objects into 3
subsamples according to their infrared luminosity (Elbaz et al.
2002).
 As a result, there are 299 ULIGs and LIGs
($L_{IR}/L_{\odot}>10^{11}$), 451 starbursts
($10^{10}<L_{IR}/L_{\odot}<10^{11}$), and 99 normal galaxies
($L_{IR}/L_{\odot}<10^{10}$) in our sample. But the number of the
ULIGs is very small (there are only 20 ULIGs in 849 objects).

We divided each of the spectral classes from emission-line ratios
(i.e. star-forming galaxies, composite galaxies, LINERs, and Seyfert
2s) further into 3 infrared luminosity bins as above. The related
numbers of the objects are given in Table~\ref{number}.

%----------------------------------------------------------figure1
%---------------------------------------------------------figure2

%-----------------------------------------------------------------table1
\begin{table*}
\caption{Number of each subsample.} \label{number} \centering
\begin{tabular}{r|r|r|r|r|c}
\hline   {method}  & \multicolumn{4}{c|}{emission-line diagram}  &{infrared luminosity bin}   \\
\hline
 {class} & star-forming & composite & LINER & Seyfert 2 &  \\
\hline
total number   &419   & 326   & 69   & 35   & 849\\
\hline
  ULIG \& LIG  &  100   &   158 &  26 &   15 &  299  \\
 starburst        & 250   &     152 &    32 &    17 & 451  \\
  normal           & 69   &      16 &    11 &     3 & 99  \\
 \hline
\end{tabular}
\end{table*}
%====================================================================
\section{Spectral synthesis}
\label{sec.3} In this section, we fit the spectral absorptions and
continua of the sample galaxies to study their stellar populations
by using the software
STARLIGHT\footnote{http://www.starlight.ufsc.br} (Cid Fernandes et
al. 2005b, see details below). We downloaded the spectra in a
one-dimensional FITS format\footnote{website:
http://das.sdss.org/DR4-cgi-bin/DAS}, and they have been
sky-subtracted, then the telluric absorption bands were removed
before the wavelength and spectrophotometric were calibrated
(Stoughton et al. 2002). We corrected the foreground Galactic
extinction using the reddening maps of Schlegel et al. (1998) and
then shifted the spectra to the rest frame. To obtain the general
view of the stellar populations of galaxies in different spectral
classes and to improve the S/N of the spectra, we combined the
spectra of all the objects to be one spectrum in each of the
spectral classes: star-forming galaxies, composite galaxies, LINERs,
and Seyfert 2s. Likewise, to check the effect of infrared luminosity
on stellar populations of galaxies, we took star-forming galaxies as
representative and combined their spectra in each of the three
infrared luminosity bins. We also did the same stellar population
synthesis fittings on the individual spectrum of the 849 sample
galaxies. The results will be discussed in Sect.4.

%============================================================
\subsection{The method}
\label{sec.3.1} The software STARLIGHT was developed by Cid
Fernandes and his colleagues to synthesize the stellar absorptions
and continua of the sample galaxies (Cid Fernandes et al. 2005b,
2007; Mateus et al. 2006,  Asari et al. 2007). It is a program to
fit an observed spectrum $O_{\lambda}$ with a model $M_{\lambda}$
that adds up to $N_{\ast}$ SSPs with different ages and
metallicities from the evolutionary synthesis models of Bruzual \&
Charlot (2003) (hereafter BC03).  The line-of-sight stellar motions
are modeled by a Gaussian distribution centered at velocity
$v_{\ast}$ and broadened by $\sigma_{\ast}$. The fit is carried out
with the Metropolis scheme (Cid Fernandes et al. 2001), which
searches for the minimum
$\chi^{2}=\Sigma_{\lambda}[(O_{\lambda}-M_{\lambda})\omega_{\lambda}]^{2}$,
where $\omega_{\lambda}^{-1}$ is the error in $O_{\lambda}$ except
for masked points. Pixels that are more than $3\sigma$ away from the
rms $O_{\lambda}-M_{\lambda}$ are given zero weight by the parameter
$'clip'$.

To extract the SSP from BC03, we used the ``Padova 1994" tracks
(Alongi et al. 1993; Bressan et al. 1993; Fagotto et al. 1994a, b;
Girardi et al. 1996)
 and a Chabrier (2003) initial
mass function (IMF). Extinction was modeled as rising from the
foreground dust, with the reddening law of Calzetti et al. (1994,
CAL hereafter) with $R_{v}=3.1$. As mentioned by the manual of
STARLIGHT that the reddening law of Cardelli et al. (1989, CCM
hereafter) was not a good choice for ULIGs, we used CAL law.
Moreover, we used a base of 45 SSPs, which included 15 different
ages from 1 Myr to 13 Gyr (i.e. 1, 3, 5, 10, 25, 40, 100, 280, 640,
900 Myr \& 1.4, 2.5, 5, 11, 13 Gyr) at each of the three
metallicities: 0.2, 1, and 2.5 $Z_{\odot}$. All bases were
normalized at $\lambda_{0}=4020\AA$. Additionally, a power law
stands for nonstellar component (Koski et al. 1978)
$F_{\nu}\varpropto\nu^{\alpha}$, was added when we fit the spectra
of AGNs (both LINERs and Seyfert 2s). We set the power-law index as
$\alpha=-1.5$ (Cid Fernandes et al. 2004b). We will first describe
the method for spectra synthesis. All spectra were sampled again in
steps of $\Delta\,{\lambda}=1\,\AA$ from 3700 to 8000 \AA, and
normalized by the median flux in the 4010 to 4060 \AA\,region. The
S/N in the normalization window spans a range of $1.1-86.1$, with
median value of 15.6. Then we compiled the mask-files for our
sample. Besides emission line and sky line, we also excluded points
that have nonzero mask
spectra\footnote{http://www.sdss.org/dr4/dm/flatFiles/spSpec.html\#specmask}
in SDSS, which are bad pixels or other artifacts. We furthermore
excluded four other windows: $5870-5905\AA$, to leap the
$Na\,D\,\lambda\lambda5890,5896$ doublet, which is from the
interstellar medium; $6845-6945\AA$ and $7550-7725\AA$, which are
listed by BC03 as their bugs due to problems in these ranges in the
STELIB library (Le Borgne et al. 2003); $7165-7210\AA$, which shows
a systematic broad residual in emission as mentioned by Mateus et
al. (2006). In addition, there were 100 objects (56 star-forming
galaxies, 34 composite galaxies, 6 LINERs, and 4 Seyfert 2s) whose
mask spectra were problematic, in which almost half of the spectra
were masked. We do not fit these 100 spectra as mentioned in
Sect.~\ref{sec.2}.

In the outputs of STARLIGHT, one of the most important parameters
to present stellar population is the population vector {\bf $x$}.
The component $x_{j} (j=1,...,N_{\ast})$ represents the fractional
contribution of the SSP with age $t_{j}$ and metallicity $Z_{j}$
to the model flux at the normalization wavelength
$\lambda_{0}=4020\AA$. Equivalently, another important parameter,
the mass fraction $\mu_{j}$, has the similar meaning.
 We analyze the obtained values of these parameters.

%======================================================================================
\subsection{The results of fitting the combined spectra according to different spectral
classes} \label{sec.3.2}

As discussed in Sect.~\ref{sec.2.1}, our sample can be divided into
four subsamples
 by a diagnostic diagram of their emission-line ratios, i.e.,
 star-forming galaxies, composite galaxies, LINERs, and Seyfert 2s.
 To clearly show their properties and to examine whether there is any difference in the
stellar populations of different spectral classes, we combine all the spectra of the
 sample galaxies in each subsample, then we synthesize these combined spectra with improved S/N.

\begin{figure*}
   \centering
   \includegraphics[height=6cm,width=6cm]{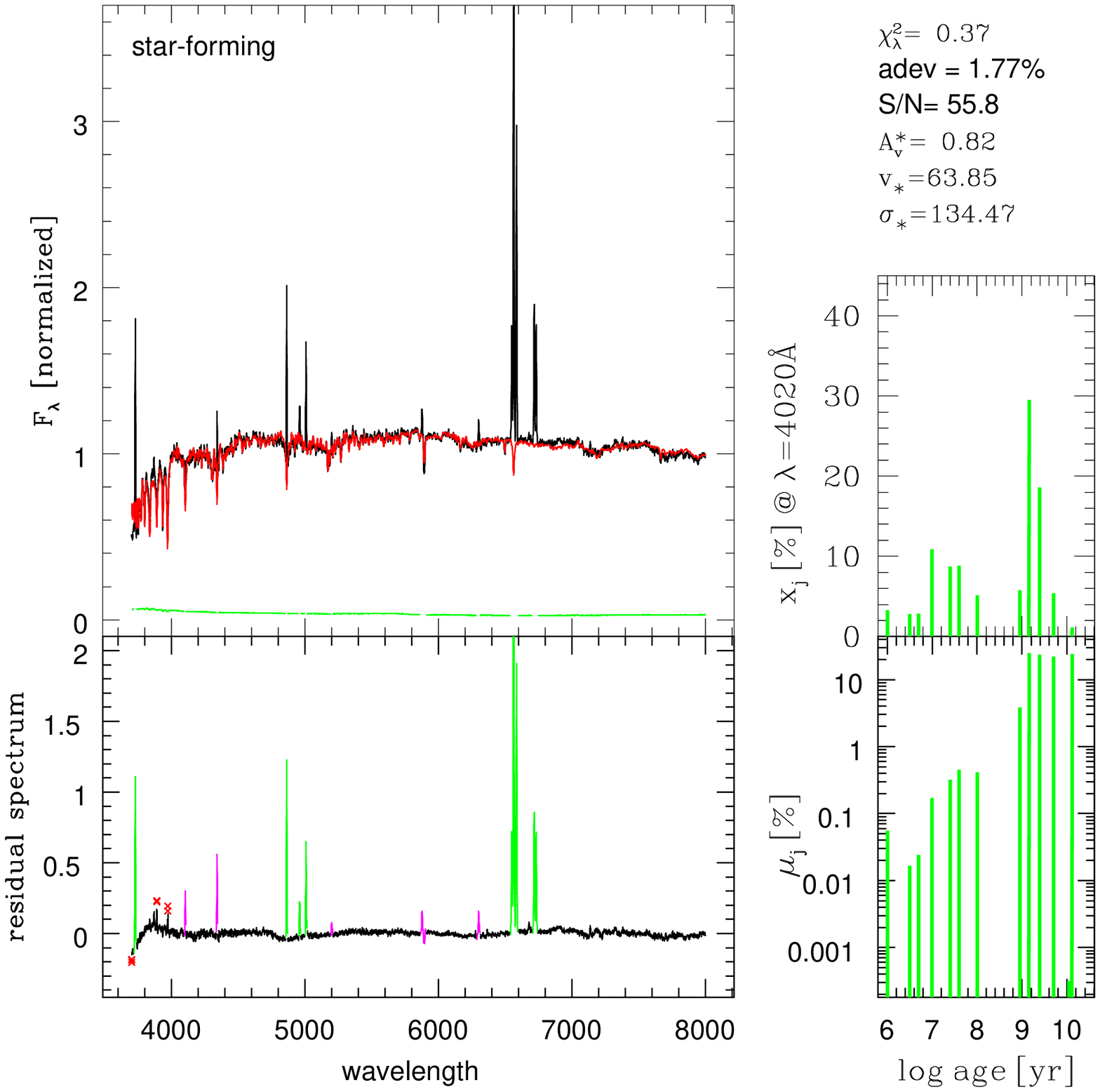}
   \includegraphics[height=6cm,width=6cm]{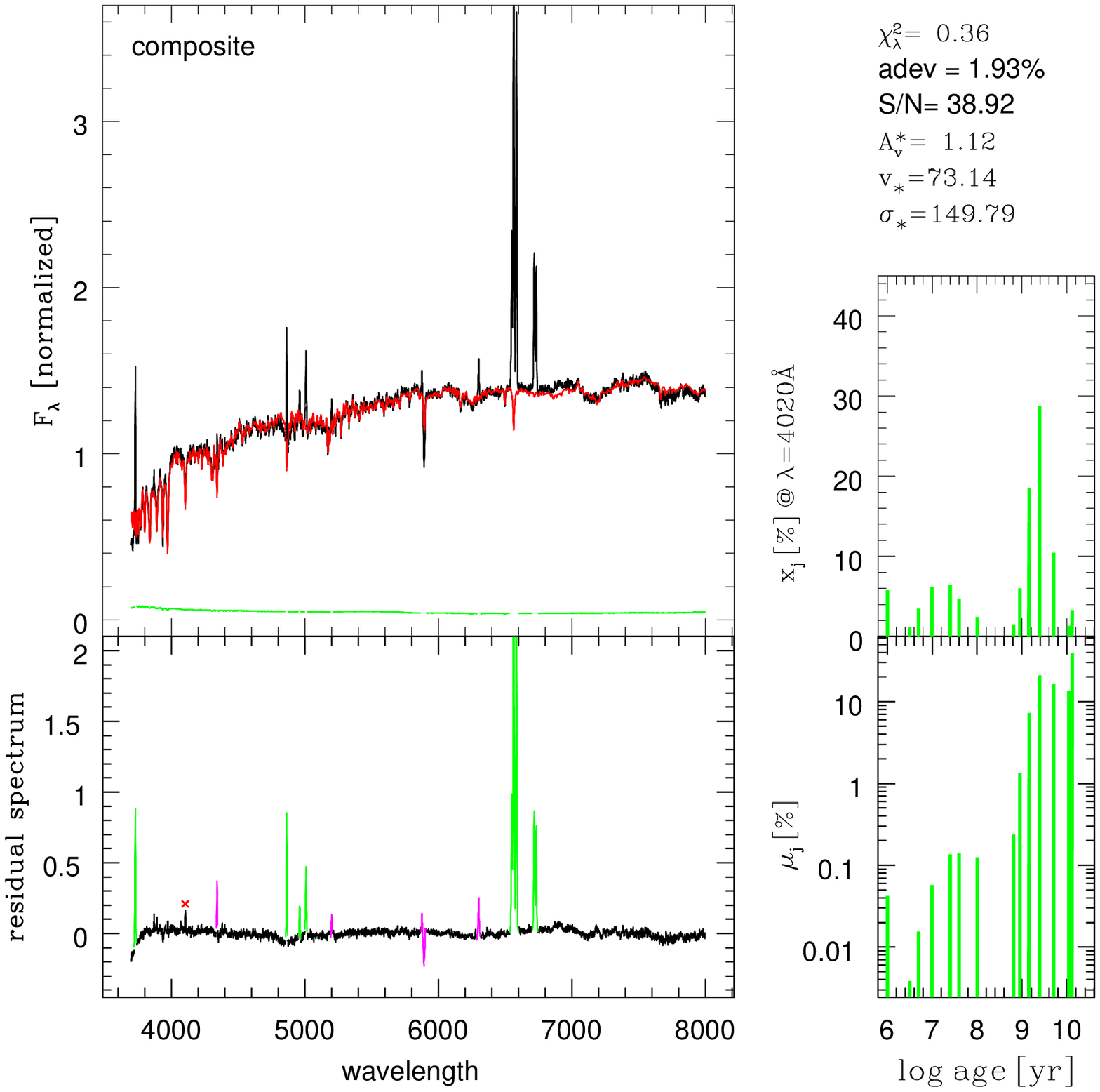}\\
   \includegraphics[height=6cm,width=6cm]{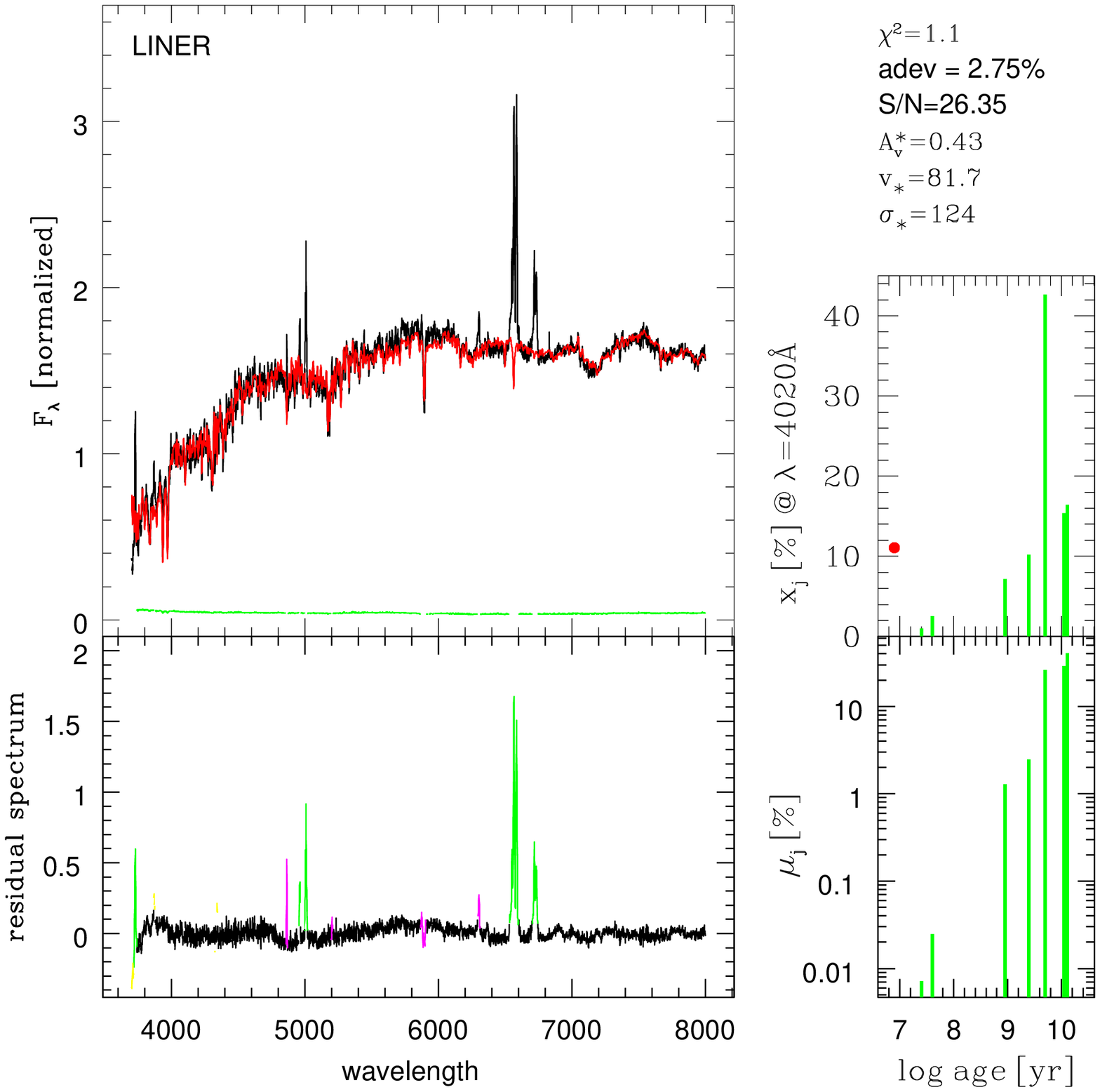}
   \includegraphics[height=6cm,width=6cm]{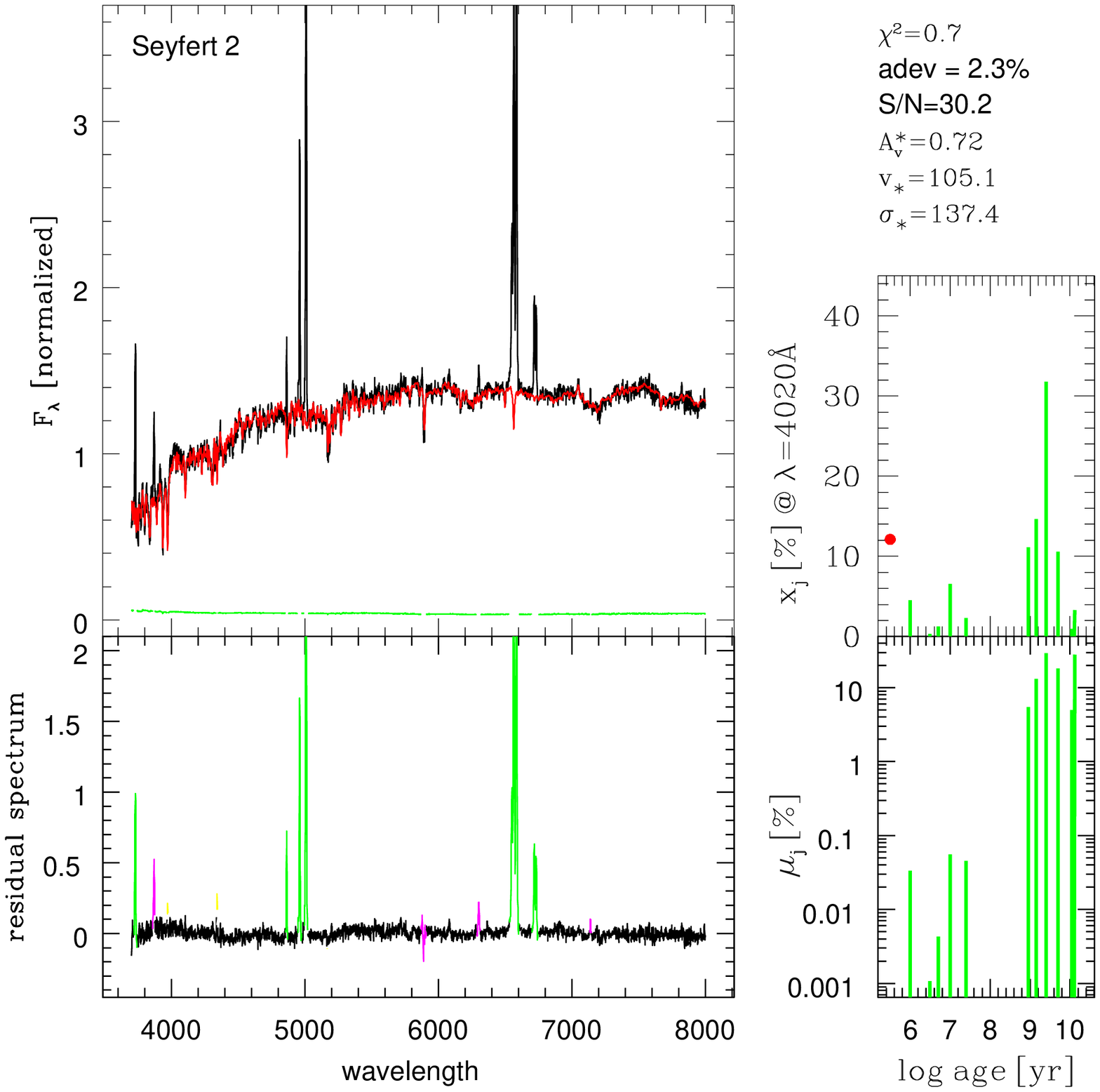}
   \caption{The spectral synthesis results of the
combined spectra of each spectral class by using 45 SSPs from BC03.
The subsamples are star-forming galaxies (top-left four panels),
composite galaxies (top-right four panels), LINERs (bottom-left four
panels), and Seyfert 2s  (bottom-right four panels). Top left: comparison of synthesis spectrum (red line)
with the observed spectrum (black line), and green line shows the
error spectrum; bottom left: the residual spectrum, green lines
represent mask regions given by SDSS flag,
 magenta lines represent mask regions given by ourselves,
and red crosses represent clip points judged by STARLIGHT code; top
right: fraction of light at $\lambda_{0}$ associated to each of the
15 ages of SSPs used in our fits. The red points in this panel of
Seyfert 2s and LINERs represent the fraction of power-law; bottom
right: fraction of mass (log-coordinate) as a function of the 15
ages of SSPs used in our fits. In the top right corner of each
group, we list some parameters explained in text (please see the
online color version for more details).} \label{combine4spectype}
\end{figure*}

Figure~\ref{combine4spectype} shows the spectral fitting results for
our combined spectra of 419 star-forming galaxies (the four panels
at top-left), 326 composite galaxies (the four panels at top-right),
69 LINERs (the four panels at bottom-left), and 35 Seyfert 2s (the
four panels at bottom-right).

%----------------------------------------------------figure3
%-----------------------------------------------------------------------------

For each of the subgroups, we plot the observed spectra, the
synthesis spectra, the error spectra, and the residual spectra (the
pure emission-line spectra) in the left two columns, and the light
and mass fractions as functions of ages of each SSPs in the right
columns. We can see that the error spectra are all very weak except
the area around $H\beta$, and the emission lines, especially the
weak ones, appear much more clearly in the residual spectra.

Additionally, at the top right of each subgroup, we also list some
other parameters, such as $\chi_{\lambda}^{2}$, i.e. the reduced
$\chi^{2}$ , the mean relative difference between synthesis and
observed spectra $\Delta_{\lambda}$; the S/N in the region of
4730-4780 \AA; the V-band stellar extinction; the velocity and the
velocity dispersion. Again we confirm that our fit is good enough
except for the area around $H\beta$. As to the bad fit around
$H\beta$, which means that there is always a trough in the continuum
around $H\beta$, Asari et al. (2007) suggests that this could relate
to calibrations in the STELIB library in this spectral range. The
good fits around H$\beta$ by using the spectra of star clusters
instead presented in Sect.~\ref{sec.3.6} confirm this. However, this
problem around H$\beta$ will not affect our general results much for
the stellar populations of the galaxies.

To obtain a general view of the stellar populations of the sample
galaxies, we arrange their stellar populations into three age bins:
old populations with age $\geq1\times10^{10}yr$, intermediate age
populations with age between $6.4\times10^{8}$ yr and
$5\times10^{9}$ yr, and young populations with age
$\leq5\times10^{8}$ yr. Then we list the contributions of each age
bin to the 4 combined spectra in percentage in Table~\ref{xj.45}.
The corresponding fractions for different infrared luminosity
bins are also given, and we show the fractions of 3 metallicity
bins. We should notice that these numbers are the percent fractions,
which is a bit different from the $x_j$ component output by
STARLIGHT directly.
 Combining Table~\ref{xj.45} and the top right panels in each
subgroup of Fig.~\ref{combine4spectype}, we find that star-forming
galaxies, composite galaxies, and Seyfert 2s contain substantial
young and intermediate age populations, while old populations give
very few contributions. As to LINERs, the dominant contributions
come from old and intermediate age populations, while there are
almost no young components. Therefore we conclude that LINERs
present the oldest stellar populations among four subsamples, and
the star-forming galaxies have more younger populations than others.
Besides that, we find the significance of young ($\leq5\times10^{8}$)
populations decreases from star-forming, composite, Seyfert 2, to
LINER. This gives credibility to our results, as Seyfert 2s and
LINERs are well known to have their activities dominated by AGNs (or
possibly shocks for LINERs), while starbursts are, in principle,
with young stars.
 These results are consistent with Boisson
et al. (2000, 2004), Kauffman et al. (2003a), Gonz\'{a}lez Delgado
et al. (2004), Cid Fernandes et al. (2008), and Stasi\'{n}aka et al.
(2008).

Another phenomenon we noticed is that the dominant contributors to
mass are all old stellar populations in four subsamples, and we
speculate that it is due to the much quicker evolution of young
massive stars than old ones. As mentioned in Sect.~\ref{sec.3.1}, we
added a featureless continuum (power-law) in fitting spectra of
LINERs and Seyfert 2s, and this component contributes almost 10\% to
the light of galaxies (Table~\ref{xj.45}). For another test, we
removed the power-law component of LINERs, making their young
populations increase about 8\%. On the other hand, if we add a power
law to composite galaxies, then the power law contributes little
(about 3\%). There are some changes, but the dominant stellar
populations are unchanged, which confirms our results.

The templates that we used in doing spectral synthesis are 45 in
total, including 15 SSPs at each of the three metallicity grids
$Z=0.2Z_{\odot}$, $Z_{\odot}$, $2.5Z_{\odot}$. This could let us
check the metallicity effect further, although the degeneracy of
metallicity and age is popularly known.  Then we obtained the
contributions of each of the three metallicity grids by summing
their contributions at different ages, which are given in
Table~\ref{xj.45}. It shows that the star-forming and composite
galaxies are dominated by the populations of $Z=0.2Z_{\odot}$,
meaning that most of their populations have metallicities even lower
than $Z_{\odot}$, and LINERs and Seyfert 2s are more metal-rich. The
populations of $Z=Z_{\odot}$ significantly contribute to LINERs,
which is consistent with the suggestion of Boisson et al. (2000,
2004), who comment that LINERs show the oldest populations in the
sample, metal-rich, with little star formation still going on. The
Seyfert 2s show a significant contribution from the SSPs with
$Z=0.2, 2.5Z_{\odot}$, which could be consistent with Cid Fernandes
et al. (2004b), who suggest that the star formation history of
Seyfert 2 nuclei is remarkably heterogeneous: young starbursts,
intermediate-age, and old stellar populations all appear in
significant and widely varying proportions.

%-----------------------------------figure4
\begin{figure*}
   \centering
   \includegraphics[height=6cm,width=6cm]{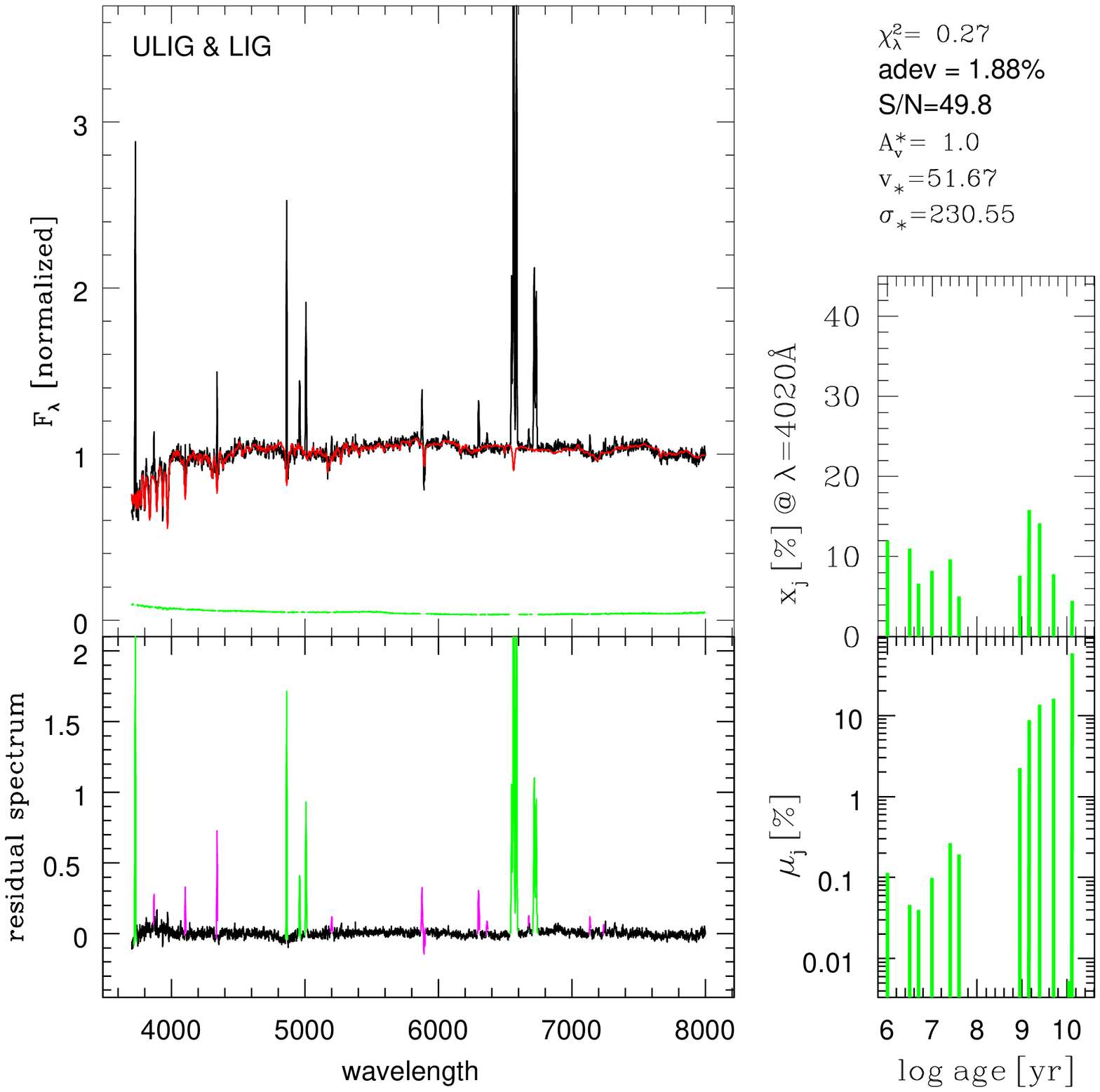}
   \includegraphics[height=6cm,width=6cm]{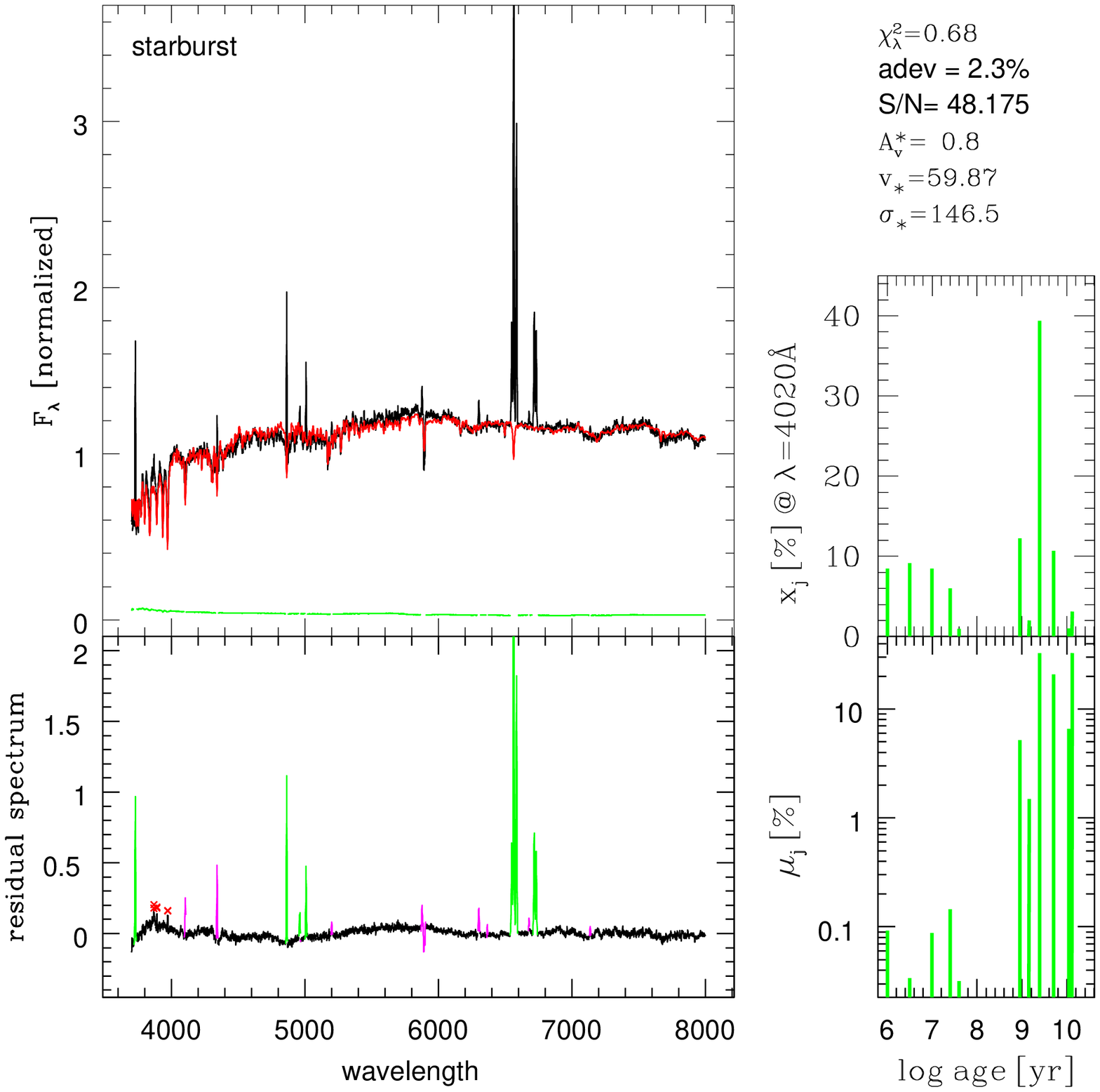}
   \includegraphics[height=6cm,width=6cm]{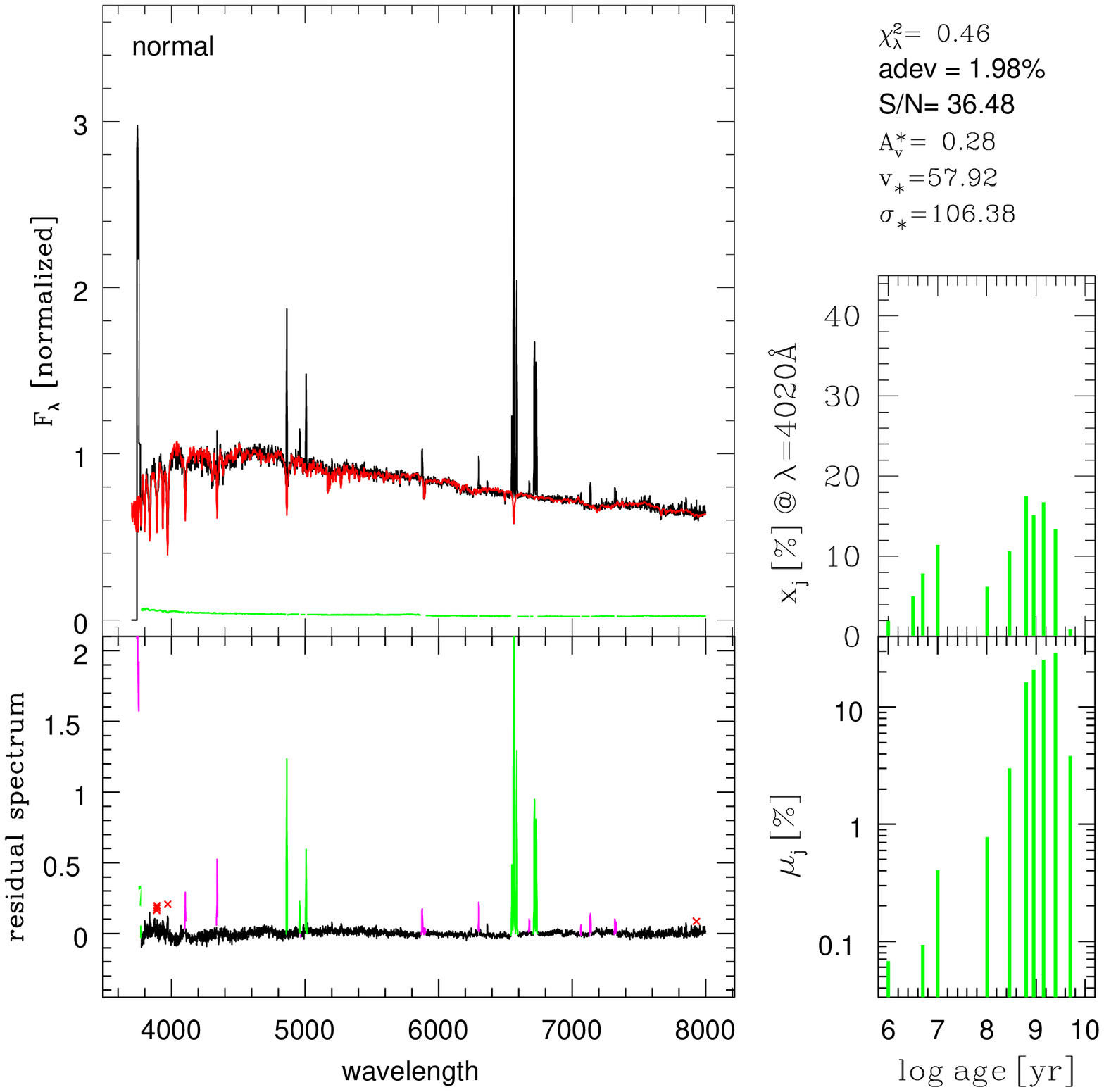}
   \caption{Spectral fittings for the combined spectra of star-forming galaxies in each
   infrared luminosity bin by using 45 SSPs from BC03. The left four panels are for ULIGs \& LIGs (
   $log(L_{IR}/L_{\odot})>11$),
 the middle four panels are for starbursts ($10<log(L_{IR}/L_{\odot})<11$),
 and the right four panels are for normal galaxies ($log(L_{IR}/L_{\odot})<10$).
 All symbols are the same as in Fig.~\ref{combine4spectype}.  }
  \label{combine3lirbin.sf}
\end{figure*}
%--------------------------------------------------------------------------table2
\begin{table*}
\caption{Stellar populations in each subsample, i. e., the results
of fitting the combined spectra by using 45 SSPs from BC03.}
\label{xj.45} \centering
\begin{tabular}{c|c|c|c|c|c|c|c|c}
\hline \multicolumn{2}{c|}{SSP}
  & \multicolumn{4}{c|}{emission-line diagram}  &
\multicolumn{3}{c}{star-forming in $L_{IR}$ bins}   \\
\hline
  &  & star-forming & composite & LINER & Seyfert 2& ULIG\&LIG & starburst & normal \\
\hline
age& young  & 41.4& 30.2& 3.4& 15.1& 51.3& 32.6& 40.4  \\
&intermediate & 57.5& 65.2& 56.4& 68.5& 44.4& 63.4& 59.6 \\
&old& 1.1& 4.6& 29.9& 4.3& 4.3& 4.0& 0.0 \\
&power law & & &  10.3& 12.1& & &  \\
\hline
$Z/Z_{\odot}$ &0.2 &78.5& 84.8& 40.4& 44.& 71.9& 76.9& 42.7\\
 &1.0 & 14.4& 10.5& 40.4& 7.4& 19.0& 15.1& 14.4 \\
 & 2.5& 7.1& 4.7& 8.9& 36.5& 9.1& 8.0& 42.9\\
&power law & & & 10.3& 12.1& & &\\
\hline
\end{tabular}
\end{table*}

%=========================================================================================
\subsection{The results of fitting the combined spectra according to infrared luminosity}
\label{sec.3.3}

To investigate whether the stellar populations of galaxies correlate
at all with the infrared luminosity and to avoid the effects by
spectral class, we specifically combine the spectra of star-forming
galaxies in three different
 infrared luminosity bins (as mentioned in Sect.~\ref{sec.2.2}),
 and fit their spectra.
We list the spectral fitting results in
Fig.~\ref{combine3lirbin.sf}.

Similar to Fig.~\ref{combine4spectype}, we get very weak error
spectra and good residual spectra in left columns  except areas
around $H\beta$. Moreover, in Table~\ref{xj.45}, we also list the
contributions of each age bin to the combined spectra. Combining
Table~\ref{xj.45} and the light fraction distributions in
Fig.~\ref{combine3lirbin.sf}, we can see that all subsamples contain
both remarkable young and intermediate age populations and a few old
populations, while ULIGs \& LIGs present the most significant
fraction of young populations among three subsamples. The population
with intermediate age is the dominant one in the starbursts and
normals. As for the stellar populations in each grid of
metallicities, the normal galaxies are more metal-rich than ULIGs \&
LIGs and starbursts. The last two are dominated by the
$Z=0.2Z_{\odot}$ components.
%========================================================================================
\subsection{ The spectral synthesis for combined spectra by using SSPs of $Z_{\odot}$ and $0.2Z_{\odot}$ }
\label{sec.3.4}

To avoid the age-metallicity degeneracy and to further check the
robustness of our estimations about the stellar populations of the
sample galaxies in different age bins, we only adopt the 15 SSPs at
each of $Z=Z_{\odot}$ and $Z=0.2Z_{\odot}$ to re-do all the fittings
on the 7 combined spectra studied above. The results are given in
Tables~\ref{comxj.15} and \ref{comxj.02}, respectively. The way
we arrange age bins are the same as before. The
corresponding fractions for different infrared luminosity bins are
also given. The dominant contributions have no obvious changes
except the $'old'$ populations increase (and the intermediate ones
decrease) by some fractions, $\sim$5-15\% in the $Z=Z_{\odot}$ case,
and $\sim$10-30\% in the $Z=0.2Z_{\odot}$ case. The corresponding
changing fractions are about 1-7\% and 2-20\% for star-forming
galaxies in infrared luminosity bins. These could be understood
simply because the old populations should increase their
contributions to supply the lack of metal-rich components.
Therefore, we believe that our estimates about the stellar
populations of the sample galaxies are robust.

%--------------------------------------------------------------------table3
\begin{table*}
\caption{Stellar populations in each subsample, i. e., results of
fitting the combined spectra by using 15 SSPs ($Z=Z_{\odot}$) from
BC03. } \label{comxj.15} \centering
\begin{tabular}{c|c|c|c|c|c|c|c|c}
\hline \multicolumn{2}{c|}{SSP}
  & \multicolumn{4}{c|}{emission-line diagram}  &
\multicolumn{3}{c}{star-forming in $L_{IR}$ bins}   \\
\hline
  &  & star-forming & composite & LINER & Seyfert 2& ULIG\&LIG & starburst & normal \\
\hline
age& young  & 38.6& 33.5& 0.1& 12.0& 50.6& 31.4& 31.8  \\
&intermediate & 55.6& 57.4& 43.4& 54.3& 39.3& 65.0& 66.9 \\
&old& 5.8& 9.1& 39.7& 19.0& 10.1& 3.6& 1.3 \\
&power law & & &  16.8& 14.7& & &  \\
\hline
\end{tabular}
\end{table*}
%---------------------------------------------------------------------table4
\begin{table*}
\caption{Stellar populations in each subsample, i. e., results of
fitting the combined spectra by using 15 SSPs ($Z=0.2Z_{\odot}$)
from BC03.}
\label{comxj.02} \centering
\begin{tabular}{c|c|c|c|c|c|c|c|c}
\hline \multicolumn{2}{c|}{SSP}
  & \multicolumn{4}{c|}{emission-line diagram}  &
  \multicolumn{3}{c}{star-forming in $L_{IR}$ bins}   \\
  \hline
    &  & star-forming & composite & LINER & Seyfert 2& ULIG\&LIG & starburst & normal \\
    \hline
    age& young  & 41.6& 35.0& 3.9& 16.1& 56.2& 37.0& 47.9  \\
    &intermediate & 45.4&43.7& 34.3& 51.1& 24.4& 57.1& 37.6 \\
    &old& 13.0& 21.3& 61.8& 25.9& 19.4& 5.9& 14.5 \\
    &power law & & &  0.0& 6.9& & &  \\
    \hline
    \end{tabular}
    \begin{list}{}{}
    \item[Notes:]All columns have the same content as Table~\ref{comxj.15}.
    \end{list}
    \end{table*}
%====================================================================
\subsection{The spectral synthesis for typical spectra by using SSPs of $Z_{\odot}$}
\label{sec.3.5}

\begin{figure*}
   \centering
   \includegraphics[height=6cm,width=6cm]{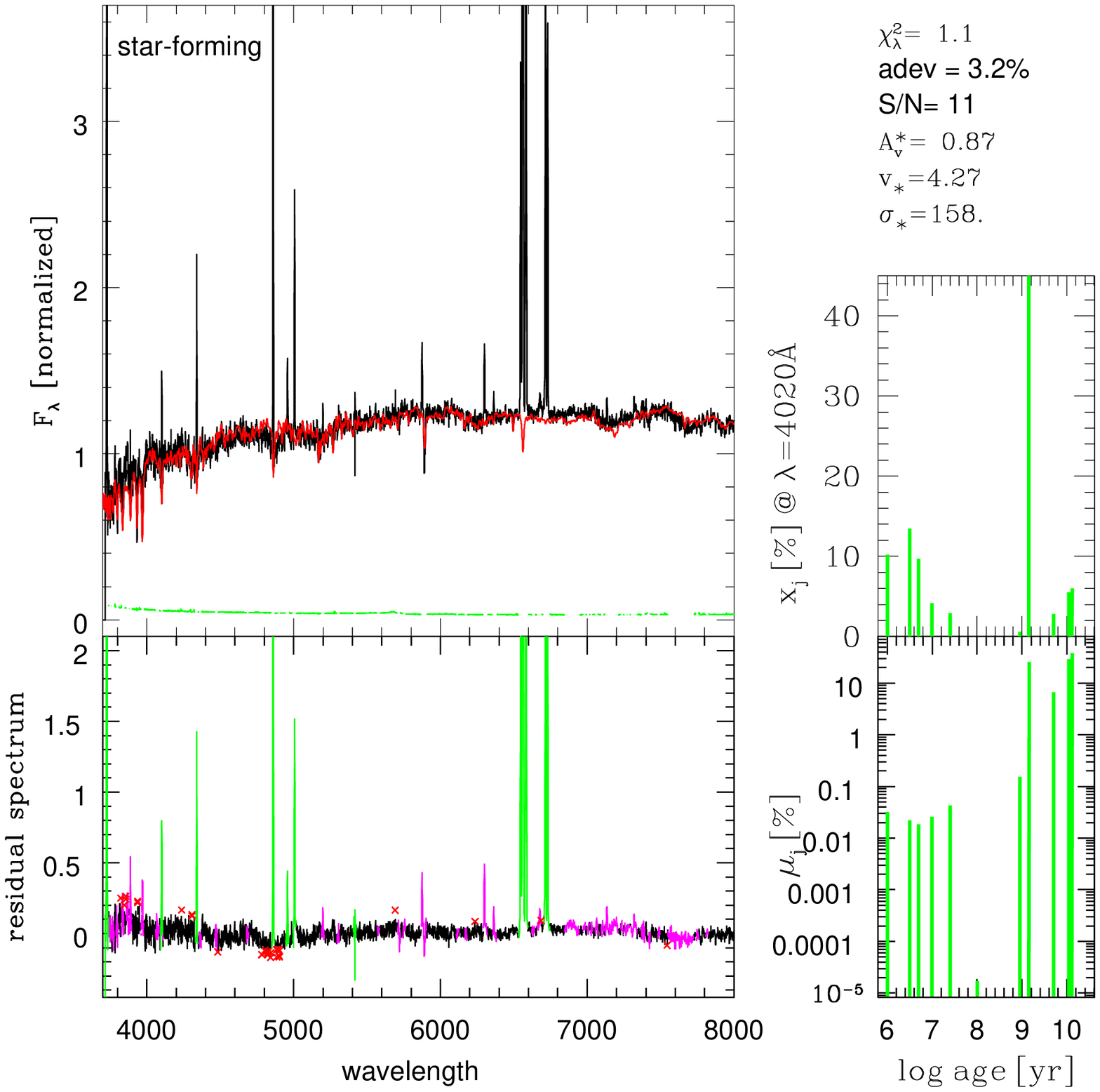}
   \includegraphics[height=6cm,width=6cm]{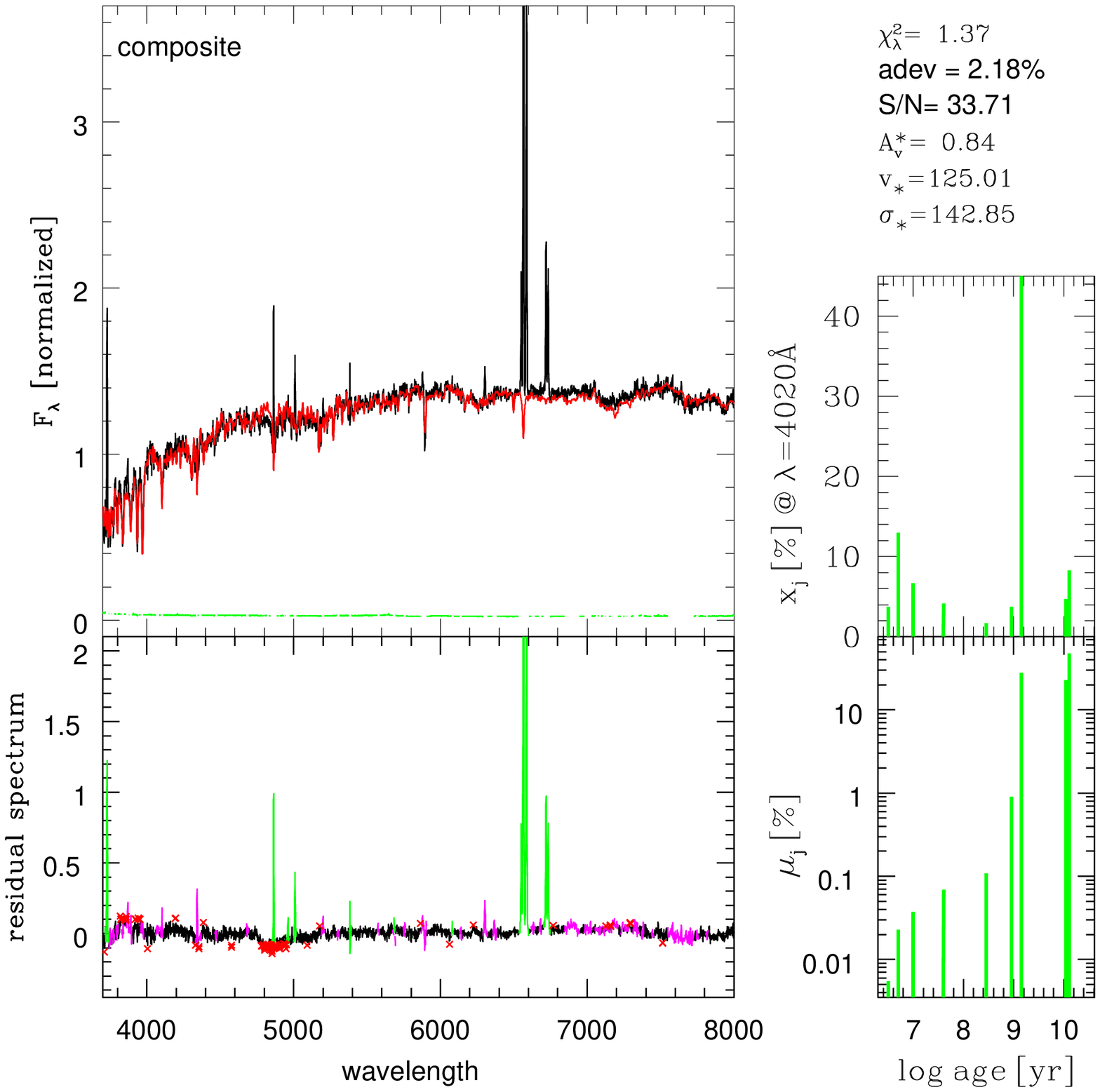}\\
   \includegraphics[height=6cm,width=6cm]{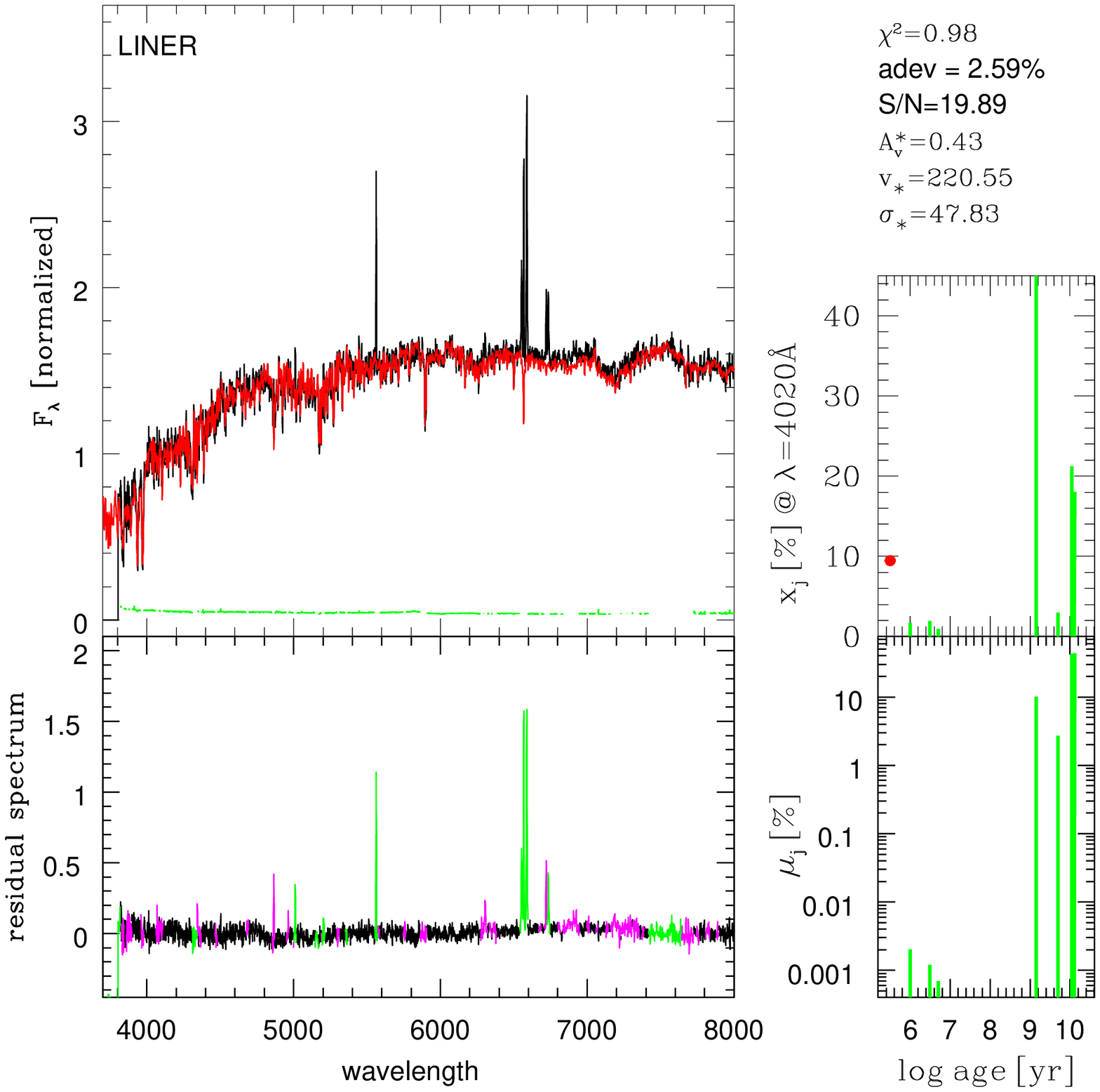}
   \includegraphics[height=6cm,width=6cm]{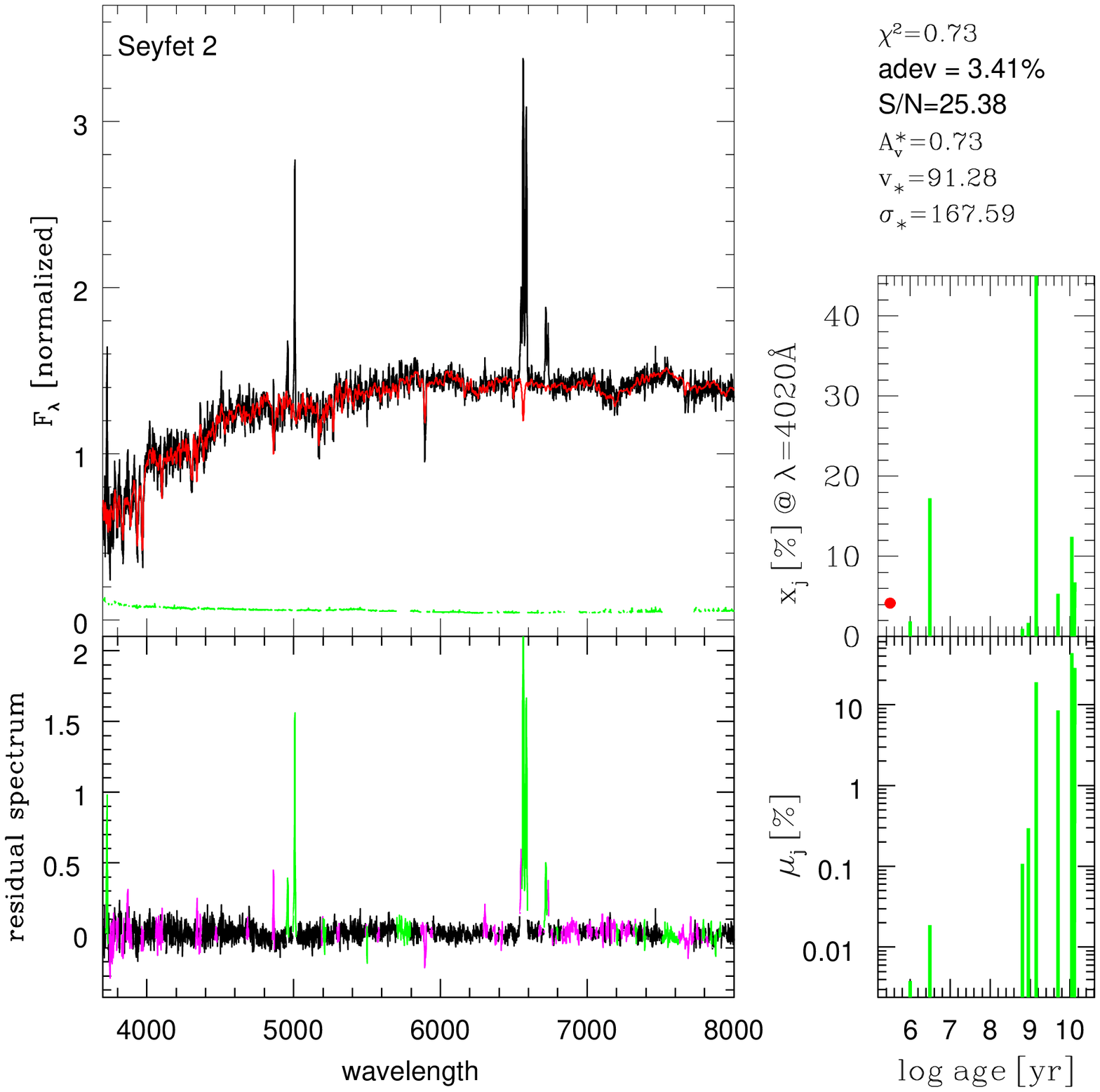}
\caption{Spectral synthesis of 4 typical spectra of 4 spectral
classes by using 15 SSPs ($Z=Z_{\odot}$) from BC03 as bases:
top-left four panels are for star-forming galaxy (PID-MJD-FID:
0310-51990-579), top-right four panels are for composite galaxy
(0499-51988-631), bottom-left four panels are for LINER
(0845-52381-633), and bottom-right four panels are for Seyfert 2
(0933-52642-185). All symbols are the same as in
Fig.~\ref{combine4spectype}.} \label{typi4class.15bc.eps}
\end{figure*}

\begin{figure*}
\centering
   \includegraphics[height=6cm,width=6cm]{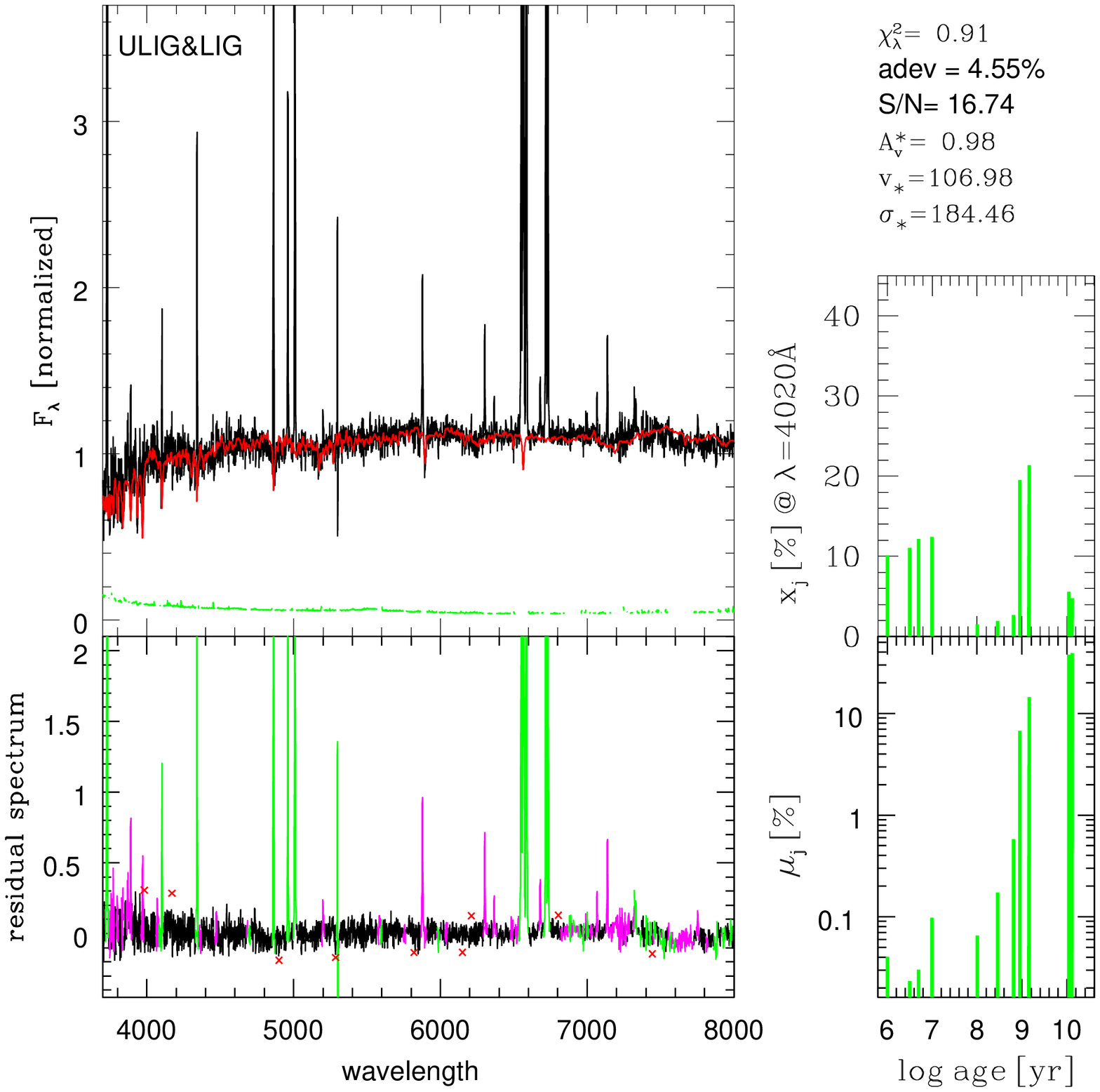}
   \includegraphics[height=6cm,width=6cm]{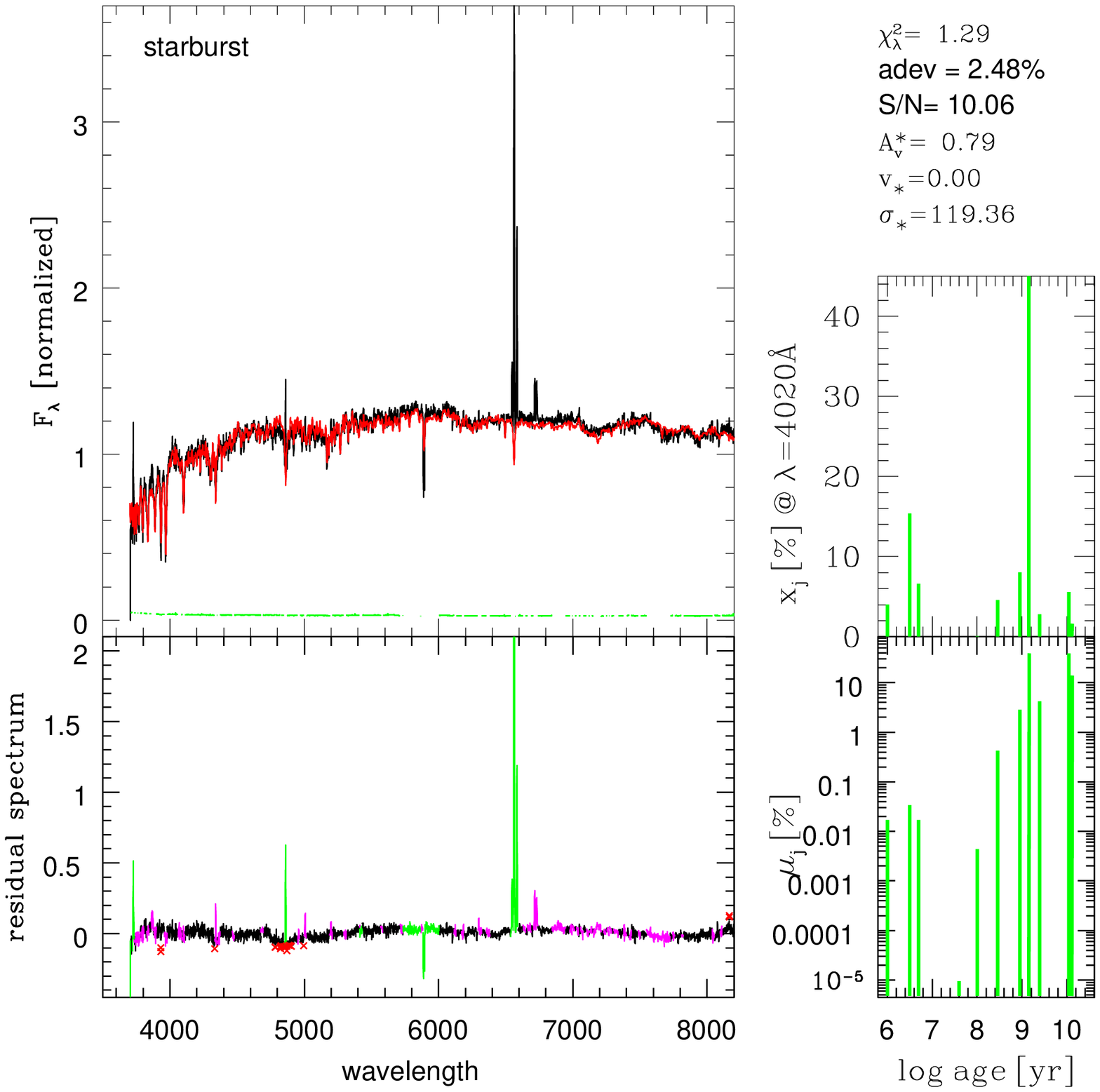}
   \includegraphics[height=6cm,width=6cm]{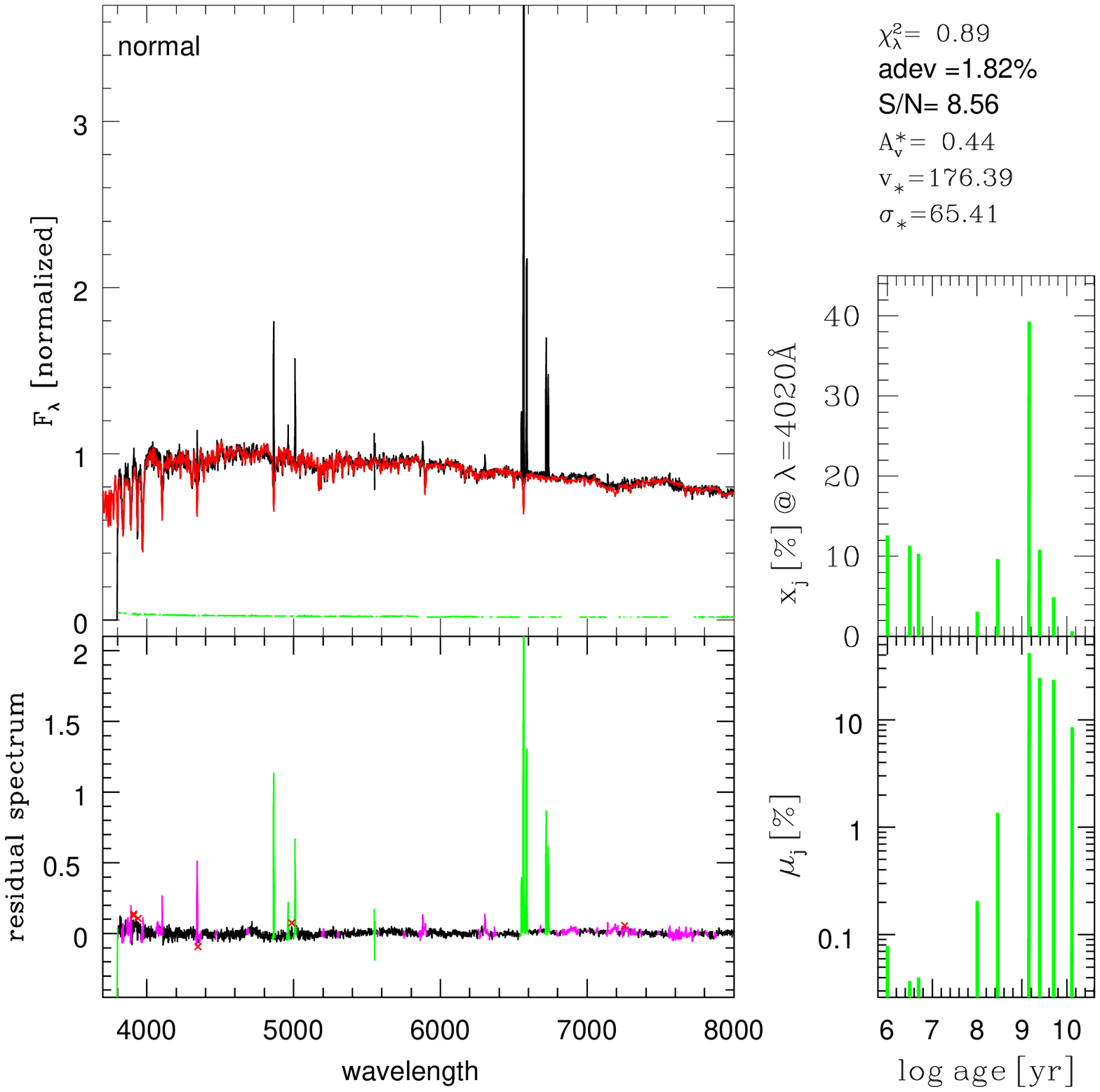}
   \caption
   {Spectral synthesis of 3 typical spectra in each $L_{IR}$ bin of
   star-forming galaxies
   by using 15 SSPs ($Z=Z_{\odot}$) from BC03 as bases:
   the left four panels are for the LIG (PID-MJD-FID: 0305-51613-093),
   the middle four panels are for the starburst (1310-53033-455) and
   the right four panels are for the normal galaxy (0586-52023-131).
   All symbols are same as Fig.~\ref{combine4spectype}.
   }
   \label{typi3lbin.15bc.eps}
    \end{figure*}

Up to now, we have been working on the combined spectra of galaxies
in the related classes. But it would be important to indicate the
variation in our results when a single spectrum instead of a
synthetic average spectral energy distribution (SED) is used. Thus,
we took one typical SED in each of the 7 classes and re-do all the
fits. This time we adopted 15 SSPs with $Z=Z_{\odot}$ following
Sect.\,3.4. We did not expect much differences from the results
obtained above. The results are given in Table~\ref{typixj.15} and
Figs.~\ref{typi4class.15bc.eps}, \ref{typi3lbin.15bc.eps}.
 It is true there are no obvious differences
from the results given in Table~\ref{comxj.15}, i.e., the case of
$Z=Z_{\odot}$ for the combined spectra. Also there is not much
difference between these and the results in Tables~\ref{xj.45} and
~\ref{comxj.02}. Again, these could confirm our robust analyses of
stellar populations of the sample galaxies.

%-----------------------------------------------figure5

%-----------------------------------------table5
\begin{table*}
\caption{Stellar populations in each of the 7 typical spectra, i.e.,
results of fitting typical spectra by using 15 SSPs ($Z=Z_{\odot}$)
from BC03. }
\label{typixj.15} \centering
\begin{tabular}{c|c|c|c|c|c|c|c|c}
\hline \multicolumn{2}{c|}{SSP}
  & \multicolumn{4}{c|}{emission-line diagram}  &
  \multicolumn{3}{c}{star-forming in $L_{IR}$ bins}   \\
  \hline
    &  & star-forming & composite & LINER & Seyfert 2& ULIG\&LIG & starburst & normal \\
    \hline
    age& young  & 39.3& 28.0& 4.3& 18.3& 47.7&29.7& 45.8  \\
    &intermediate & 49.6& 59.5& 50.0& 59.3& 42.2& 63.3& 53.5 \\
    &old& 11.1& 12.5& 36.9& 18.4&10.1 & 7.0& 0.7 \\
    &power law & & &  8.8& 4.0& & &  \\
    \hline
    \end{tabular}
\begin{list}{}{}
\item[Notes:] All columns have the same content as Table~\ref{comxj.15}. 
\end{list}
\end{table*}

%----------------------------------------------figure6

%======================================================================
\subsection{The spectral synthesis by using spectra of star clusters}
\label{sec.3.6}
\begin{figure*}
   \centering
   \includegraphics[height=6cm,width=6cm]{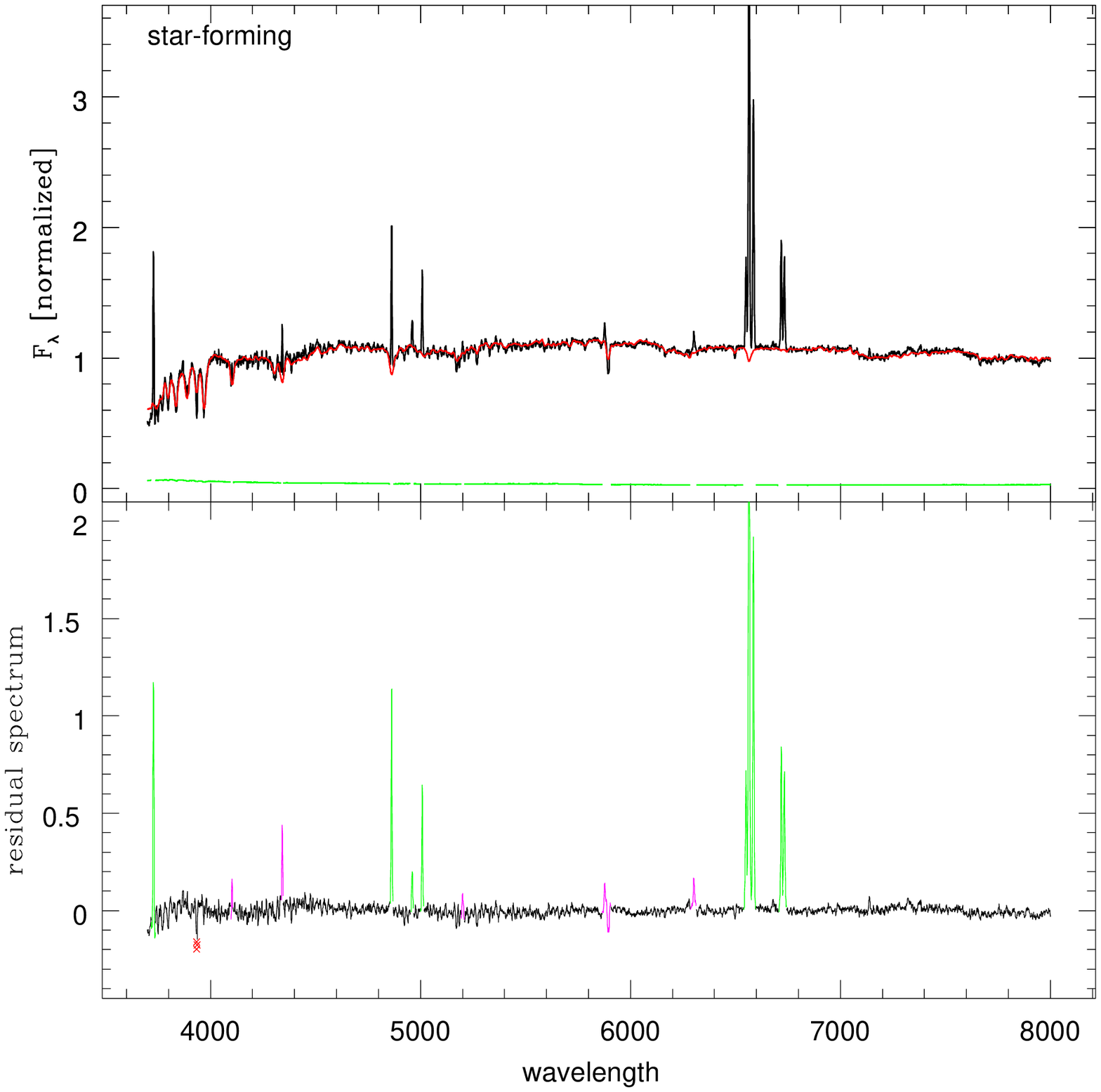}
   \includegraphics[height=6cm,width=6cm]{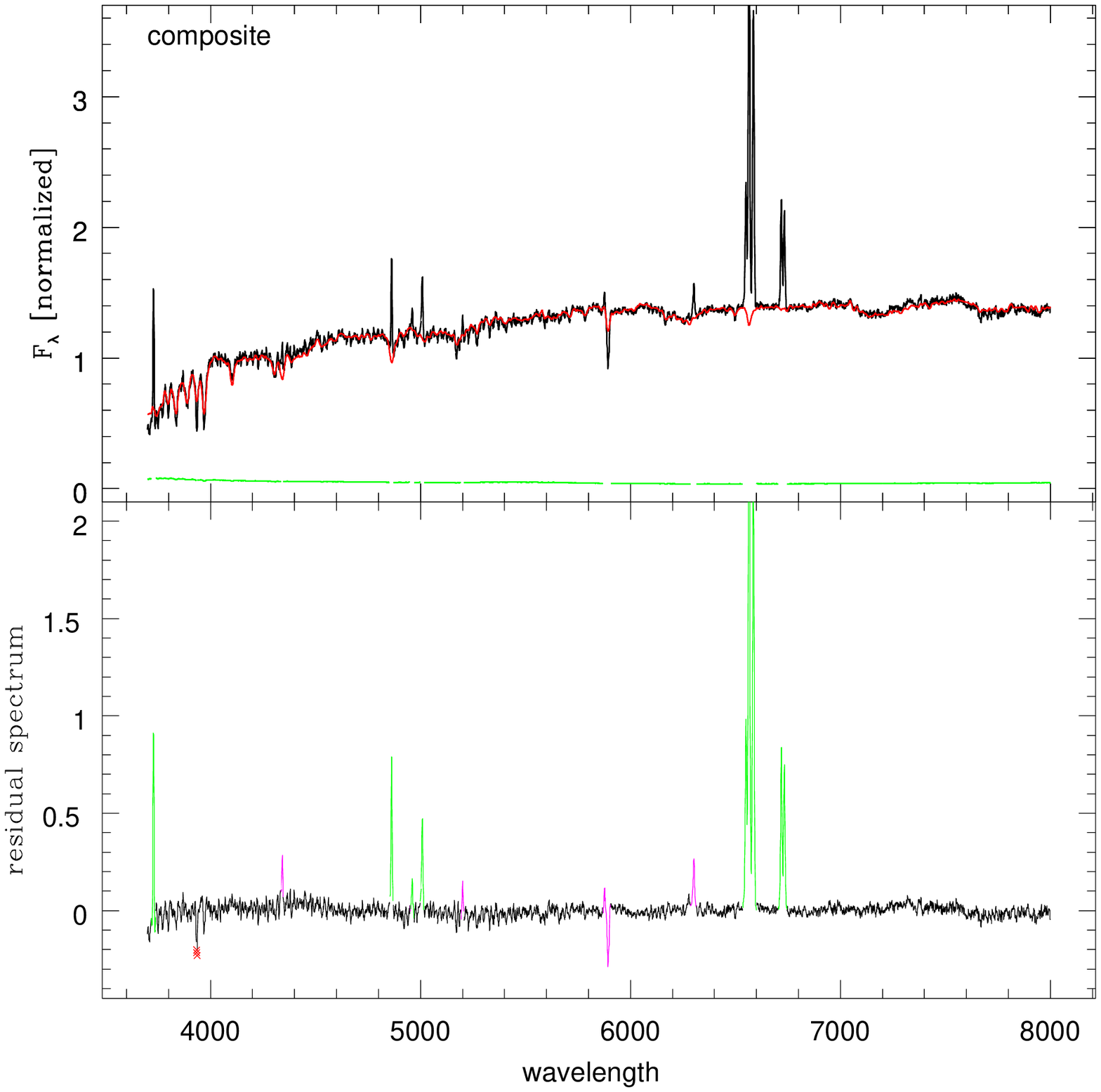}\\
   \includegraphics[height=6cm,width=6cm]{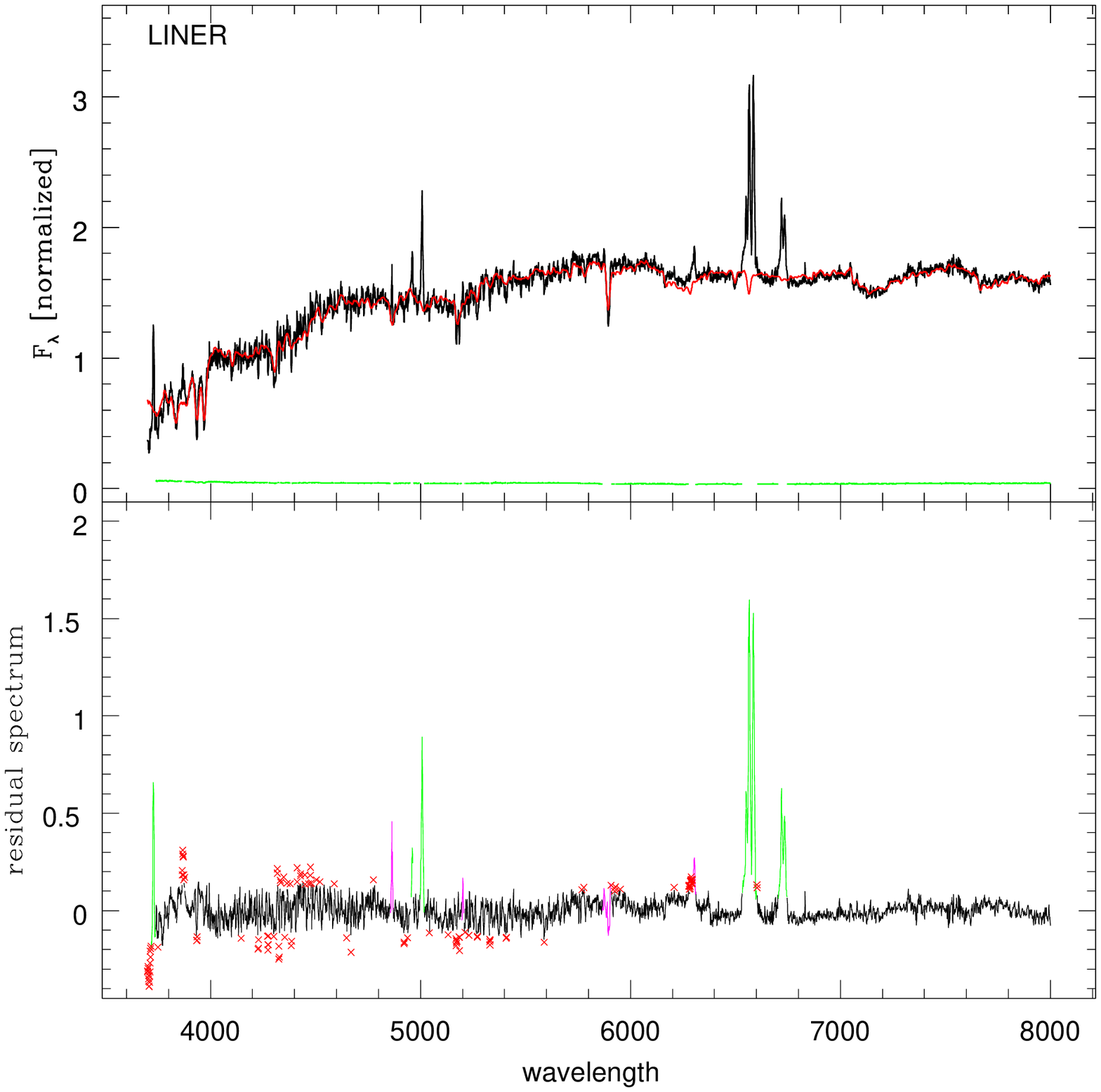}
   \includegraphics[height=6cm,width=6cm]{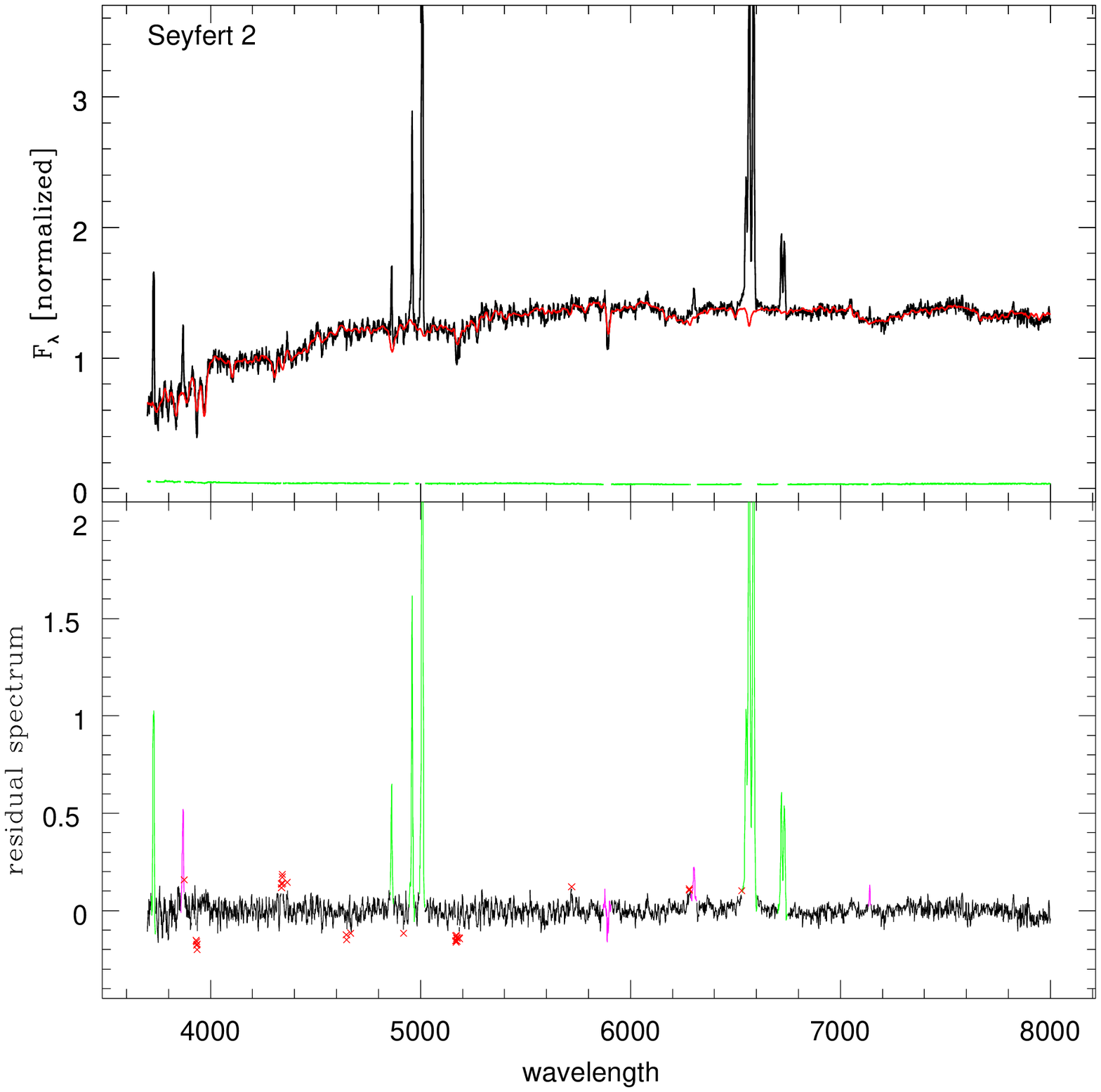}
\caption {Spectral synthesis of the combined spectra of 4 spectral
classes by using the spectra of star clusters as bases: top-left two
panels are for star-forming galaxies, top-right two panels are for
composite galaxies, bottom-left two panels are for LINERs, and the
bottom-right two panels are for Seyfert 2s. All symbols are the same
as in the left hand columns of Fig.~\ref{combine4spectype}. }
\label{sc4combclass.eps}
\end{figure*}

\begin{figure*}
\centering
   \includegraphics[height=6cm,width=6cm]{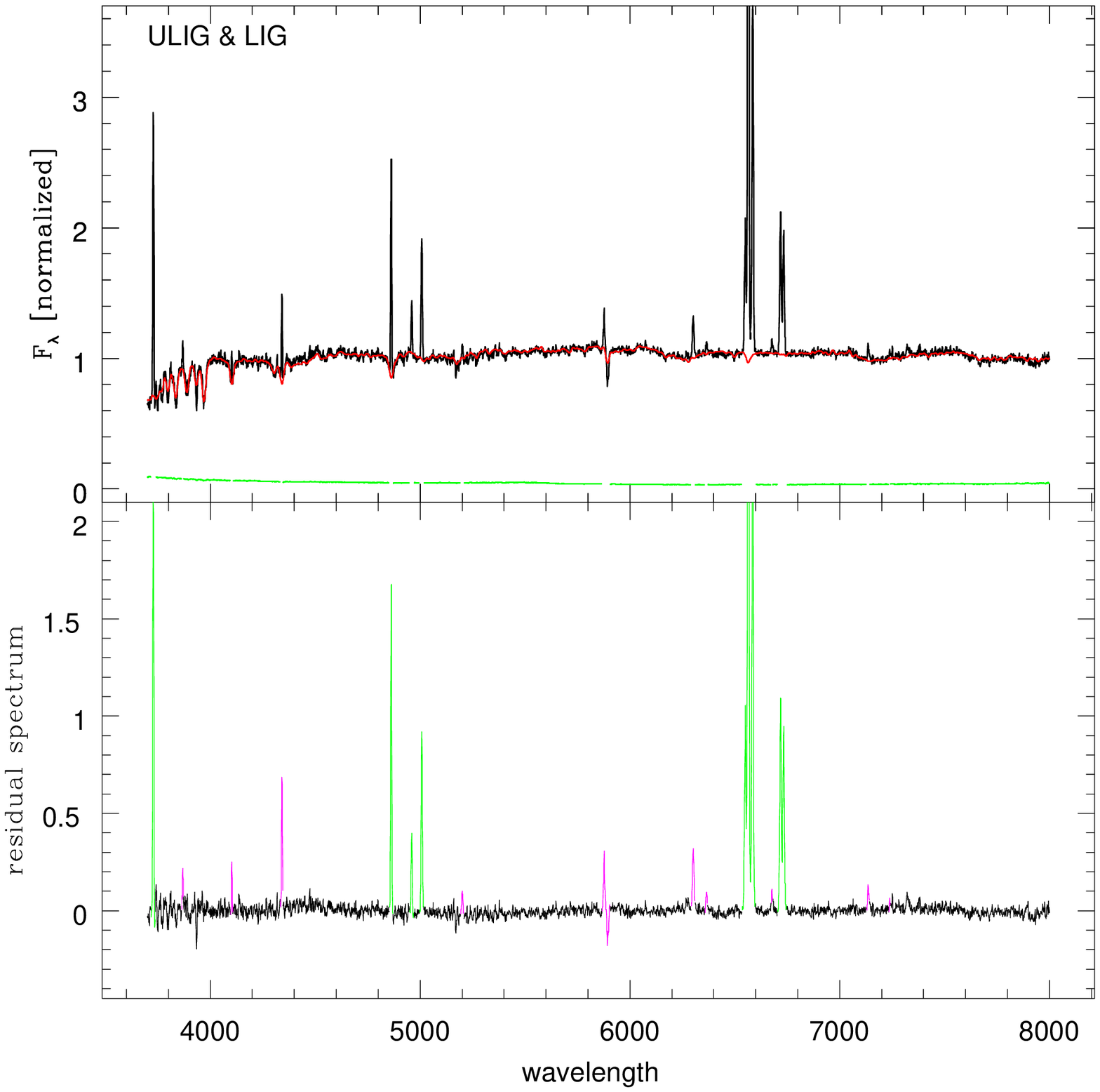}
   \includegraphics[height=6cm,width=6cm]{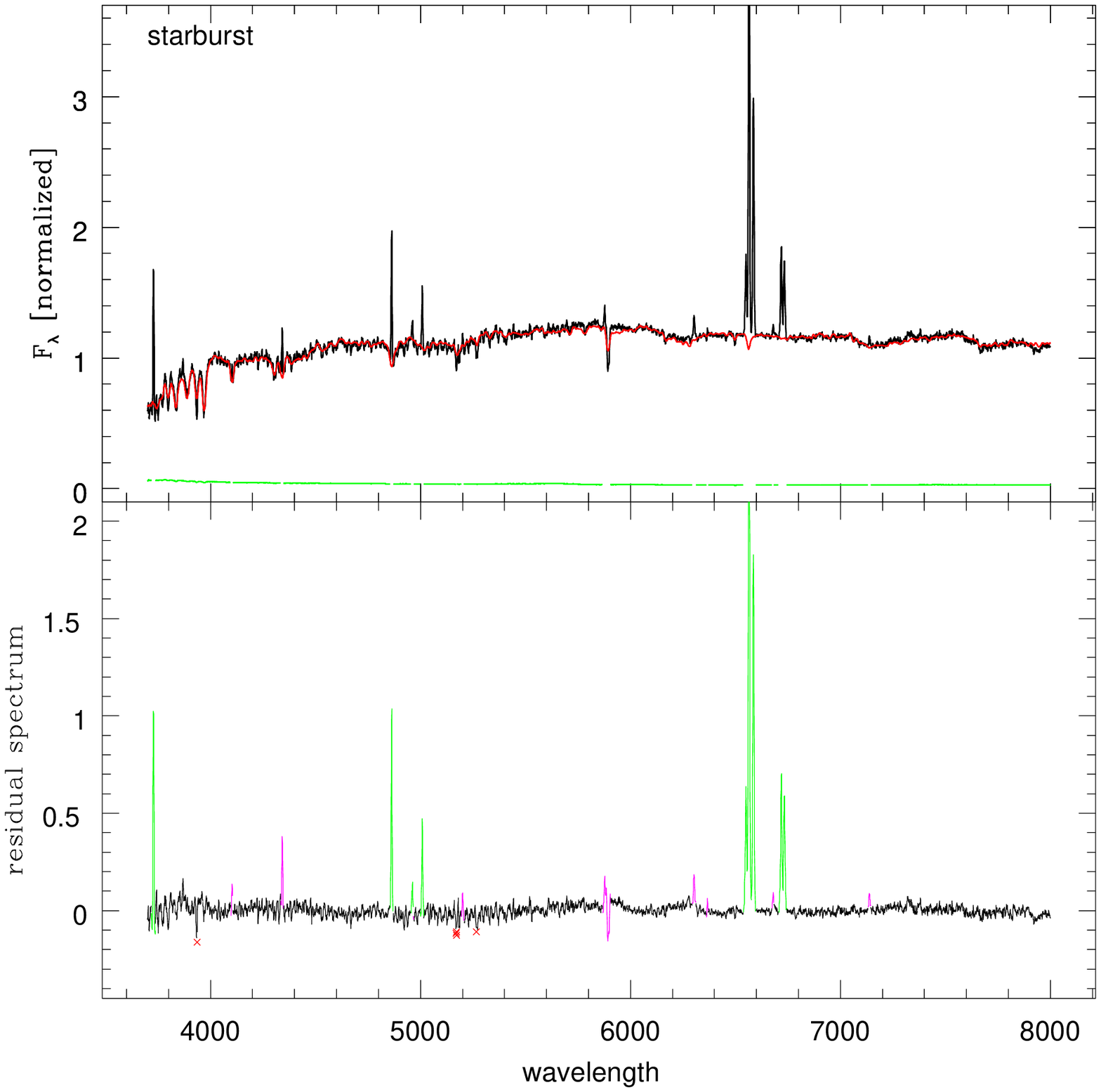}
   \includegraphics[height=6cm,width=6cm]{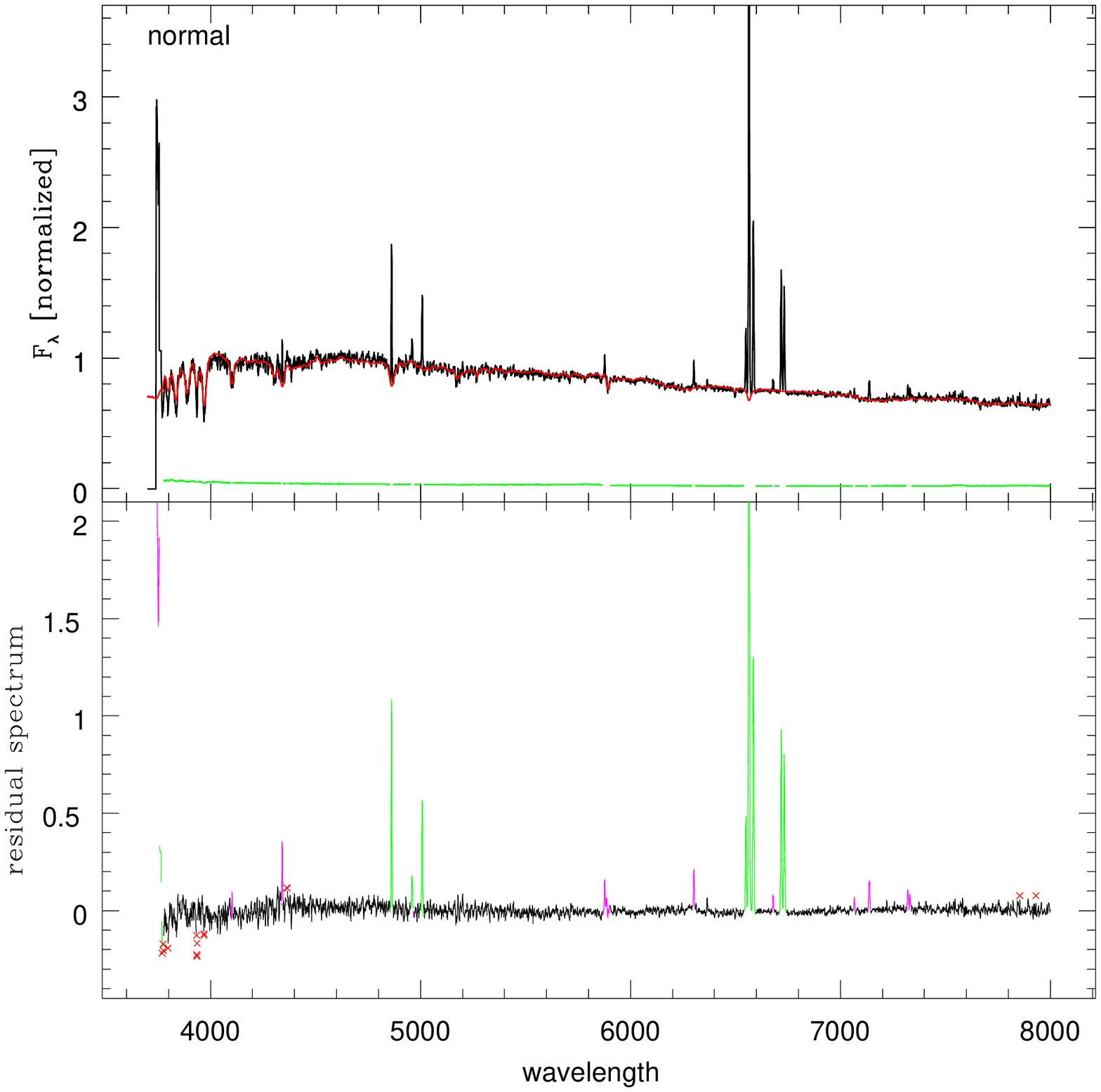}
   \caption
   {
   Spectral synthesis of 3 combined spectra of star-forming galaxies in each
   $L_{IR}$ bin by using the spectra of star clusters as bases:
   the left two panels are for ULIGs \& LIGs, the middle two panels are for
   starbursts,
   and the right two panels are for normal galaxies. All symbols are same
   as in
   the left hand columns of Fig.~\ref{combine4spectype}.
           }
   \label{sc3comblbin.eps}
    \end{figure*}

In this section, we use 15 spectra of star clusters with different
ages and metallicities given in Bica \& Alloin
 (1986a,b) to re-fit the combined spectra of the subgroup
 galaxies. The resolutions of the two sets of spectra have been matched.
 The results are given in
 Figs.~\ref{sc4combclass.eps} and \ref{sc3comblbin.eps} for the
 different spectral classes and infrared luminosity bins.
We can see that after changing
 our bases into spectra of the star cluster, the area around $H\beta$
 obtains a much better fit.

We then arranged all bases into 4 age-bins: old populations 
$\geq 10^{10}yr$, young populations $\leq 5\times10^{8}yr$,
intermediate age populations $5\times10^{8}-10^{10}yr$,
and H\,II. We list the stellar population fractions of each age bin
in all subsamples in Table~\ref{xj.15}. The same, these numbers
are the percent fractions.
 We can see that LINERs
present an overwhelming fraction from old populations, the
star-forming galaxies have younger populations than the others, and
the significance of young ($\leq5\times10^{8}$) populations
decreases from star-forming, composite, Seyfert 2, to LINER. These
are consistent with the results obtained by using the BC03's SSP. As
to the star-forming galaxies in different $L_{IR}$ bins, the results
are consistent with those from the BC03's SSP as well.

In addition, we also used the SSPs of Vazdekis/Miles (Vazdekis 1999,
S\'{a}nchez-Bl\'{a}zquez et al. 2006, Koleva et al. 2008)
 instead of those of BC03
to re-fit the spectra, and the results are consistent with those
from the BC03's SSP.

%----------------------------------------------------------------figure7
%----------------------------------------------------------------figure8
%-----------------------------------------table6
\begin{table*}
\caption{Stellar populations in each subsample, i. e., results of
fitting the combined spectra by using spectra of 15 star clusters as
bases.} \label{xj.15} \centering
\begin{tabular}{c|c|c|c|c|c|c|c}
\hline age
  & \multicolumn{4}{c|}{emission-line diagram}  &
\multicolumn{3}{c}{star-forming in $L_{IR}$ bins}   \\
\hline
  & star-forming & composite & LINER & Seyfert 2& ULIG\&LIG & starburst & normal \\
\hline
young & 40.7&28.1& 11.2& 21.0& 43.6& 25.2& 49.8 \\
intermediate &19.2& 35.5& 0.0& 21.7& 16.1& 21.4& 0.0 \\
old&40.1& 36.4& 88.8& 52.4& 32.8& 53.4& 35.0  \\
H\,II & 0.0& 0.0& 0.0& 4.8& 7.5& 0.0& 15.2\\
\hline
\end{tabular}
\end{table*}

%========================================================================
\section{The relations of some property parameters from emission lines, absorption lines, and continua }
\label{sec.4}

In this section, we investigate the relations of some property
parameters obtained from spectral synthesis and the emission lines
or continua measurements.

The outputs of STARLIGHT are divers, and we discuss some of them in
this paper, i.e., the V-band stellar extinction $A_{V}^{\ast}$, the
light-weighted mean stellar age
$\langle\,log\,t_{\ast}\rangle\,_{L}$ (Eq.~\ref{tl}), and the
mass-weighted mean stellar metallicity
$\langle\,Z_{\ast}\rangle\,_{M}$ (Eq.~\ref{zm}):

\begin{equation}
\label{tl}
\rm\langle\,log\,t_{\ast}\rangle\,_{L}=\sum_{j=1}^{N_{\ast}}x_{j}log\,t_{\ast\,
,j},
\end{equation}

\begin{equation}
\label{zm} \rm\langle\,Z_{\ast}\rangle\,_{M}=\sum
_{j=1}^{N_{\ast}}\mu_{j}Z_{\ast\, ,j},
\end{equation}
where the subscript $L$ or $M$ denotes a light-weighted or a
mass-weighted average.

The MPA/JHU collaboration has put the measurements of emission lines
and some property parameters for the SDSS galaxies on the MPA SDSS
website\footnote{http://www.mpa-garching.mpg.de/SDSS/} (Kauffmann et
al. 2003b; Brinchmann et al. 2004; Tremonti et al. 2004 etc.). These
values were obtained from the stellar-feature subtracted spectra
with the spectral population synthesis code of Bruzual \& Charlot
(2003). We used these emission-line flux measurements to obtain some
of their property parameters, or we adopted some parameters from
their catalog, e.g., $D_{n}(4000)$ and $H\delta_{A}$.
%---------------------------------------------------figure9
   \begin{figure*}
   \centering
   \includegraphics[height=6cm,width=6cm]{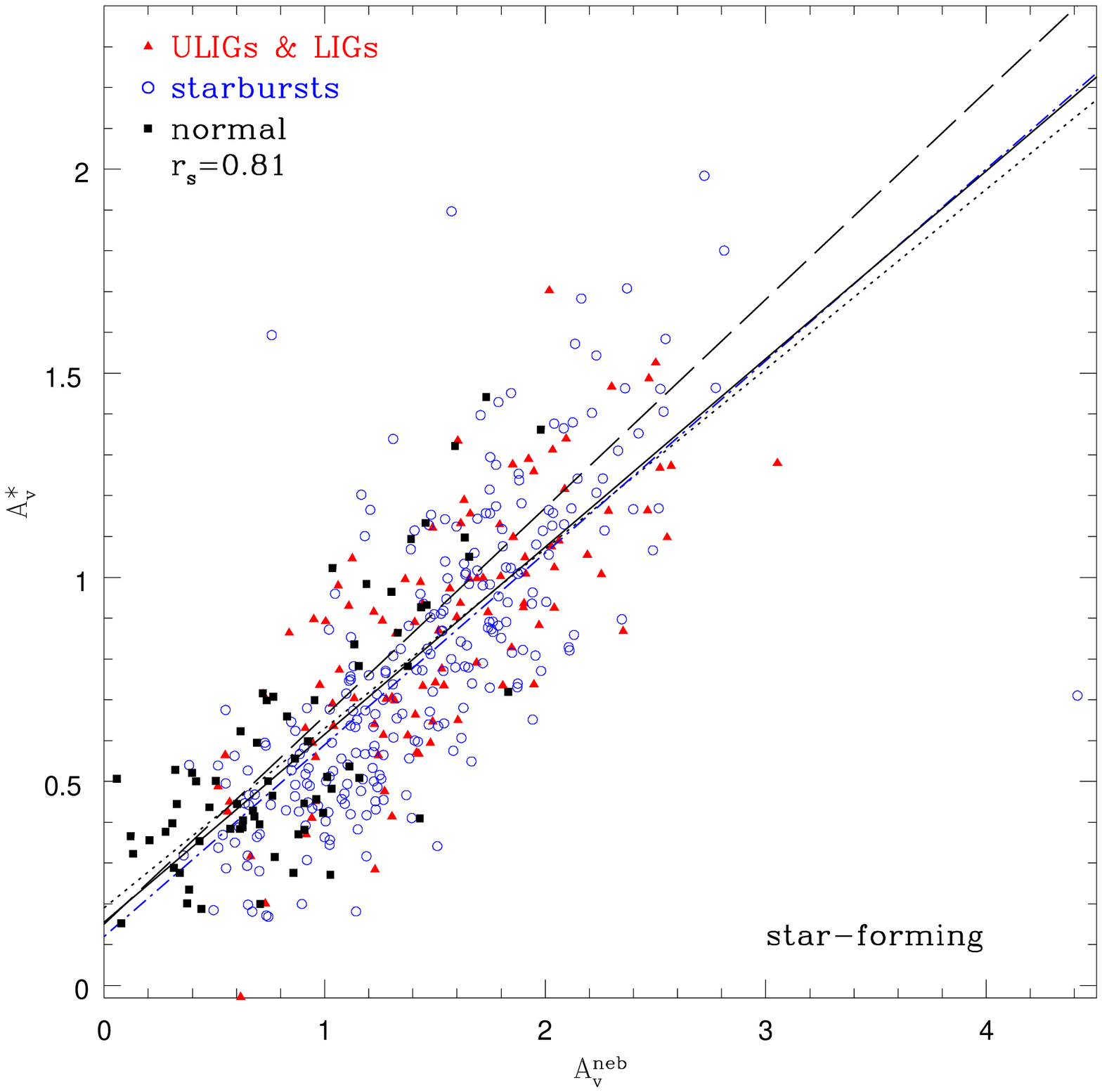}
   \includegraphics[height=6cm,width=6cm]{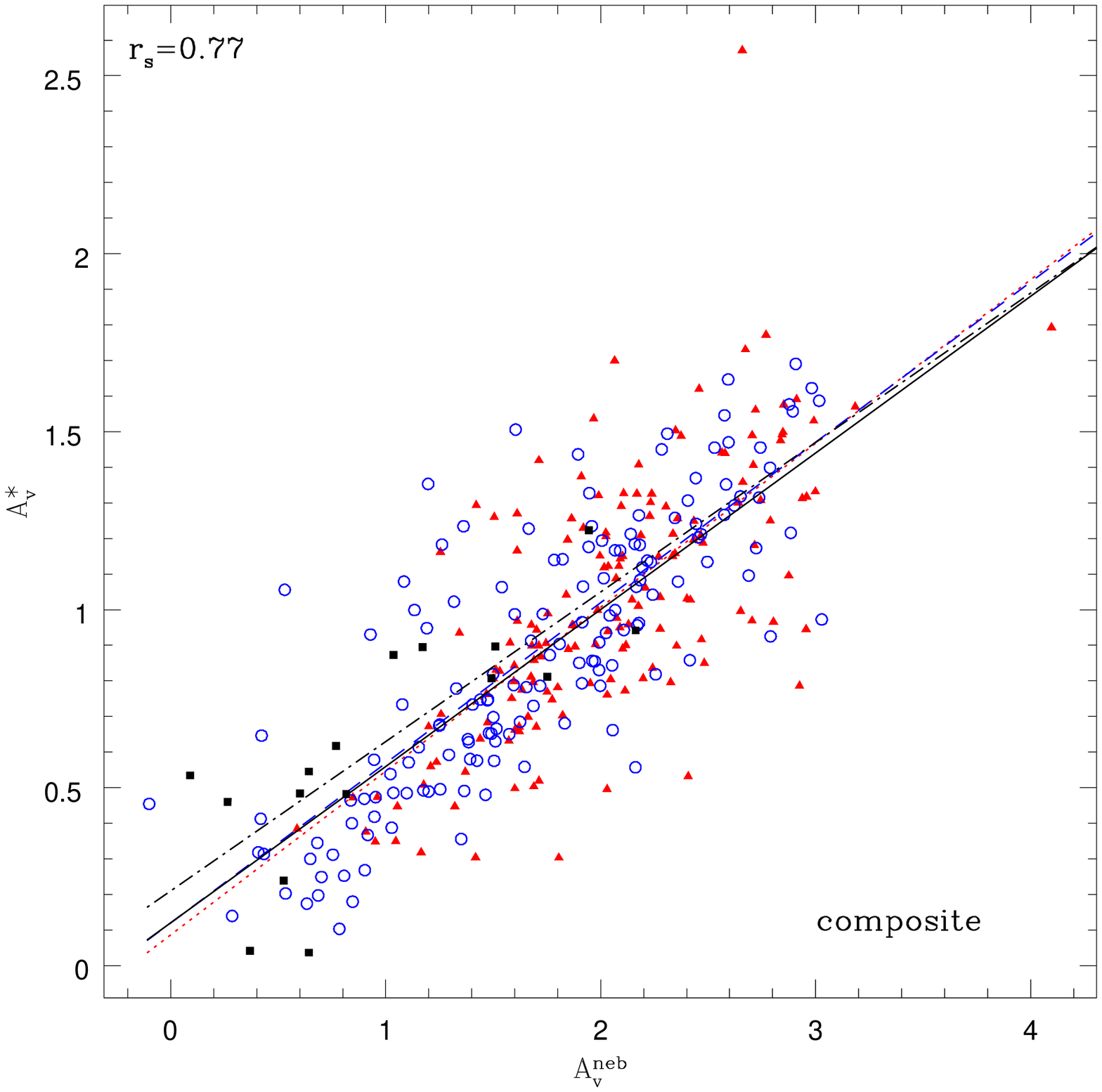}
   \includegraphics[height=6cm,width=6cm]{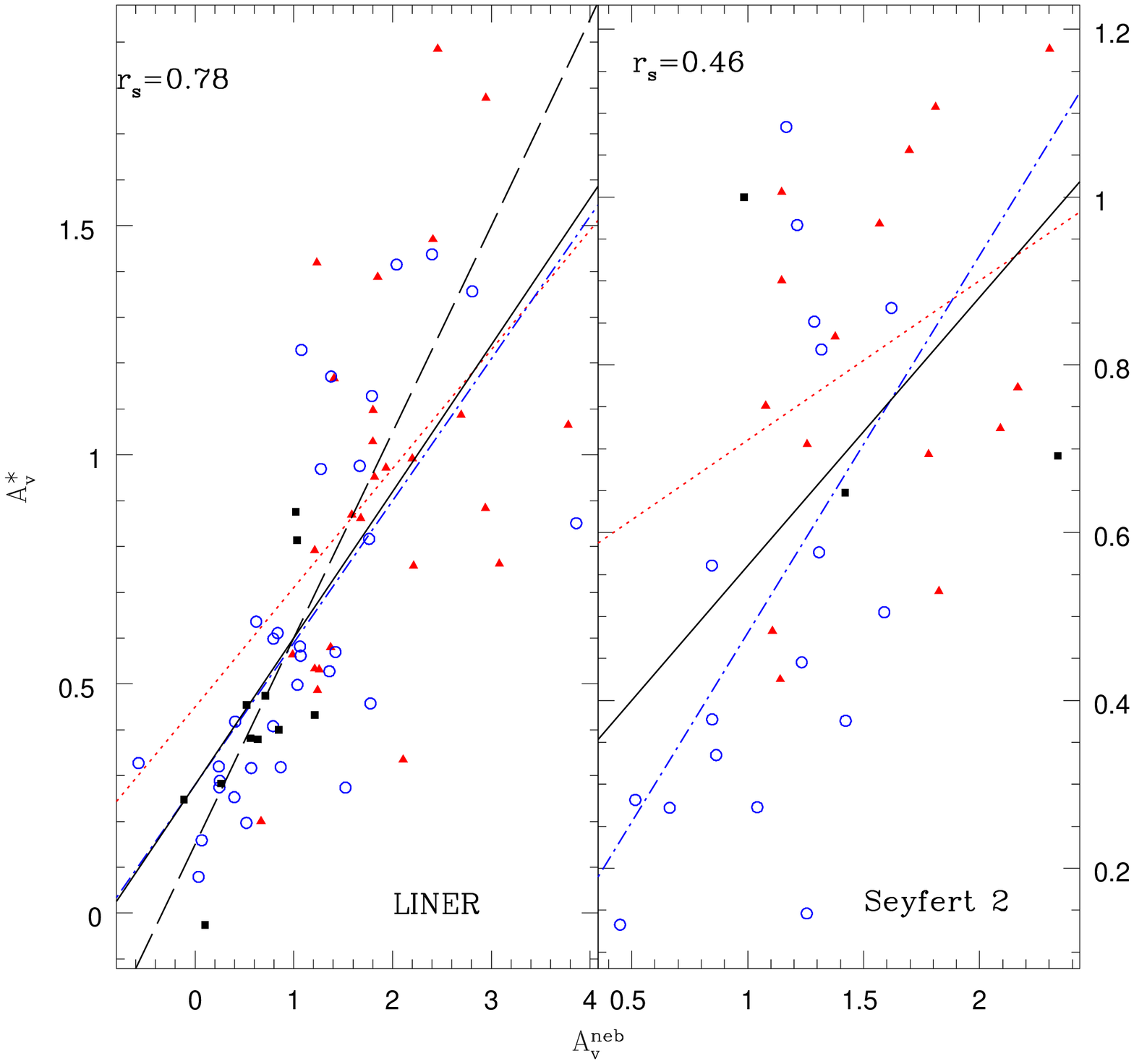}
   \caption{Comparison between stellar and nebular extinctions for
star-forming galaxies (left panel), composite galaxies (middle
panel), LINERs, and Seyfert 2s (right two panels).
 The red triangles represent ULIGs \& LIGs (i.e.
$L_{IR}/L_{\odot}>10^{11}$), blue open circles represent starbursts
(i.e. $10^{11}>L_{IR}/L_{\odot}>10^{10}$), while black filled
squares represent normal galaxies (i.e. $L_{IR}/L_{\odot}<10^{10}$).
The solid line represents a linear fit including all the sample
galaxies in the plot, the dotted line represents a linear fit only
including ULIGs \& LIGs, the dot-dashed line represents a linear fit
only including starbursts, and the long-dashed line represents a
linear fit only including normal galaxies. The Spearman-Rank Order
correlation coefficient is shown in the top left corner of each
panel for all the sample galaxies there.
          }
   \label{avneb_avstar.eps}
    \end{figure*}

%==================================================================
\subsection{Relations between stellar extinction and nebular extinction}
\label{sec.4-1}

We estimated the nebular extinction inside the galaxies from the
Balmer emission-lines ratios H$\alpha$/H$\beta$, assuming case B
recombination, with a density of 100 cm$^{-3}$ and a temperature of
$10^4$ K (Osterbrock 1989) following the relation of

\begin{equation}
\label{ext1} (\frac{I_{H\alpha}}{I_{H\beta}})_{obs}=
(\frac{I_{H\alpha0}}{I_{H\beta0}})_{intr}$10$^{-c(f(H\alpha)-
f(H\beta))}.
\end{equation}

Using the average interstellar extinction law given by Osterbrock
(1989), we obtained $f(H\alpha)-f(H\beta)=-0.37$. We then calculated
$A_{V}^{neb}$ according to Eqs.~\ref{ext2}-\ref{ext3} (Osterbrock
1989, Seaton 1979):

\begin{equation}
\rm \log(H\alpha/H\beta)_{ins}
=\log(H\alpha/H\beta)_{obs}+c\times(-0.37) \label{ext2},
\end{equation}

\begin{equation}
\rm A_{v}^{neb}=c\times 3.1/1.47 \label{ext3},
\end{equation}
where $c$ is the extinction coefficient, and we set the intrinsic ratio
$(H\alpha/H\beta)_{ins}=2.86$ for star-forming galaxies and
composite galaxies, and $(H\alpha/H\beta)_{ins}=3.1$ for AGNs
(Veilleux et al. 1995).

Comparisons between stellar extinction $A_{V}^{\ast}$ from our
spectral synthesis and nebular extinction $A_{V}^{neb}$ calculated
by using the emission-line flux measurements given in the MPA/JHU
database for each subsample are presented in
Fig.~\ref{avneb_avstar.eps}. It shows that $A_{V}^{\ast}$ and
$A_{V}^{neb}$ are strongly correlated, and the correlation
coefficients are very large.

The left hand panel of Fig.~\ref{avneb_avstar.eps} presents the two
$A_V$ values of star-forming galaxies. The linear fit including all
star-forming galaxies yields $A_{V}^{\ast}=0.15+0.46\times
A_{V}^{neb}$, which shows that the stellar extinction is almost half
of the nebular extinction. This confirms the results from other
authors (Stasi\'{n}ska et al. 2004; Cid Fernandes et al. 2005b;
Mateus et al. 2006) and accords with studies of nearby star-forming
galaxies (Calzetti et al. 1994). To check whether the correlation
between $A_{V}^{\ast}$ and $A_{V}^{neb}$ is affected by infrared
luminosity $L_{IR}$, we assigned different symbols to the 3
different $L_{IR}$ bins, and we also plotted the corresponding
linear least-square fits of them (ULIGs \& LIGs:
$A_{V}^{\ast}=0.19+0.44\times A_{V}^{neb}$, starbursts:
$A_{V}^{\ast}=0.12+0.47\times A_{V}^{neb}$, and the normal galaxies:
$A_{V}^{\ast}=0.15+0.51\times A_{V}^{neb}$). We can see that these
lines are close to the solid line (slope $\pm 0.02$) except for the
long-dashed line. Given the distributions of stellar and nebular
extinctions in the figure, we infer that the ionized gas extinction
(nebular extinction) may be smaller in normal galaxies than in
others, so that we get a steeper fit result for them.

Moreover, in the middle panel of Fig.~\ref{avneb_avstar.eps}, the
linear fit including all composite galaxies yields
$A_{V}^{\ast}=0.12+0.44\times A_{V}^{neb}$, which infers that
nebular extinction is almost twice as much as stellar extinction.
This conclusion is similar to that of star-forming galaxies.
Moreover, we mark different symbols for different bins of $L_{IR}$
and plot the linear fit lines for these subsamples. We find that all
of the three lines are close to the solid line, which is for the
total sample.

Finally, in the right two panels of Fig.~\ref{avneb_avstar.eps}, we
perform the relations between stellar extinction and nebular
extinction for LINERs (left panel) and Seyfert 2s (right panel). In
the left panel, the linear fit containing all LINERs yields
$A_{V}^{\ast}=0.28+0.32\times A_{V}^{neb}$, which suggests that the
ionized gas exceeds twice as much extinction as stars. It is
different from the results of star-forming and composite galaxies.
Furthermore, we plot three other linear fit lines for the objects in
each of the 3
 $L_{IR}$ bins in this panel, and we find that
the behavior of normal galaxies is similar to normal galaxies in
subsamples of star-forming and composite galaxies, while the other
two subsamples perform differently. In the right hand panel, Seyfert
2s show much weaker relationship between stellar and nebular
extinctions, and the Spearman-Rank Order correlation coefficient is
smaller than others. Moreover, the linear fit including all 35
Seyfert 2s yields $A_{V}^{\ast}=0.24+0.32\times A_{V}^{neb}$. As the
number of data points in Seyfert 2s' sub-sample of normal galaxies
is extremely poor, we did not try linear fit to this subsample.
There are very few points in AGNs, especially in Seyfert 2s, so we
cannot draw definite conclusion. In total, the ionized gas tends to
suffer excess extinction twice as much as stars.
%------------------------------------------------figure10
  \begin{figure*}
   \centering
   \includegraphics[height=6cm,width=6cm]{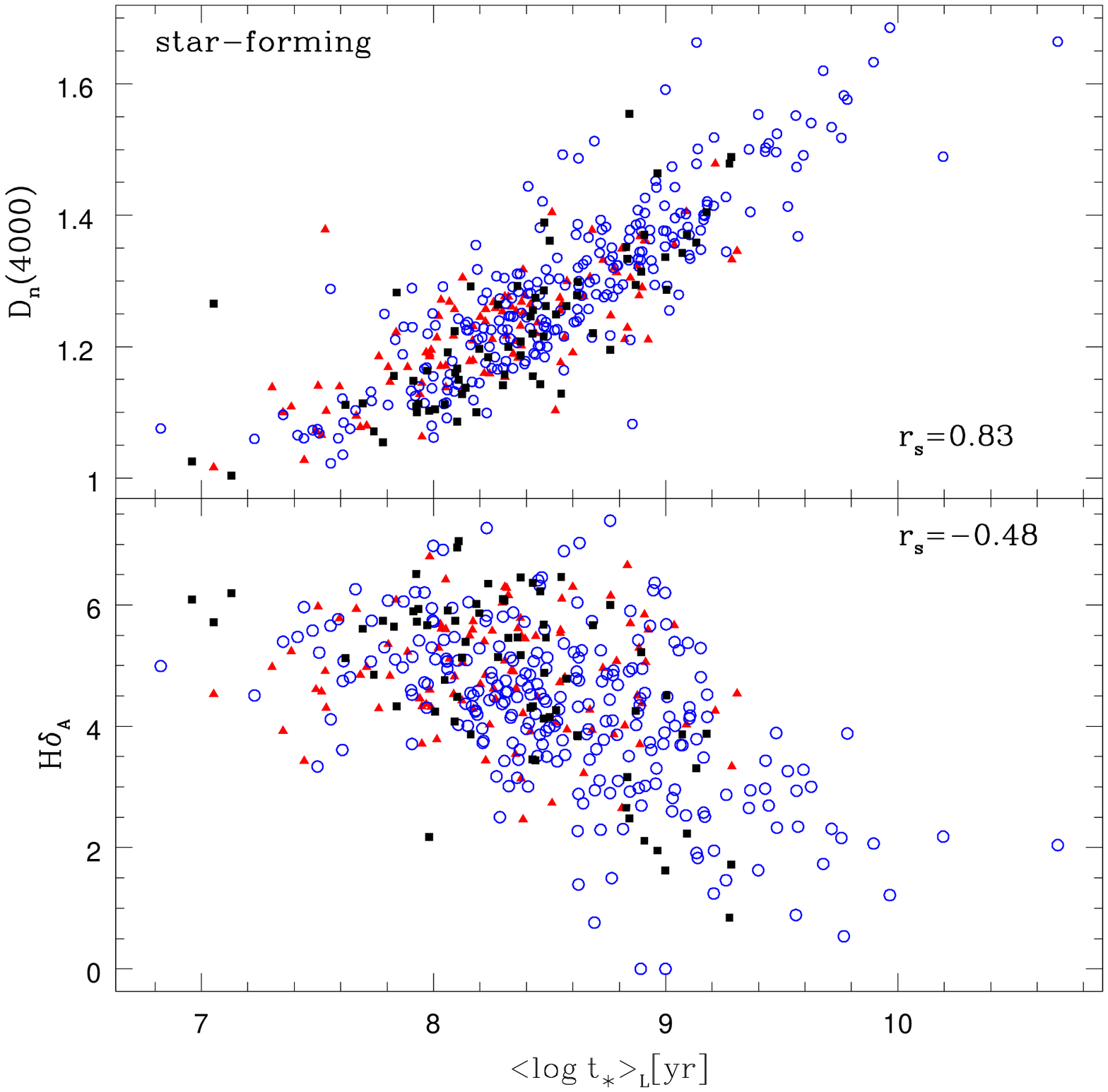}
   \includegraphics[height=6cm,width=6cm]{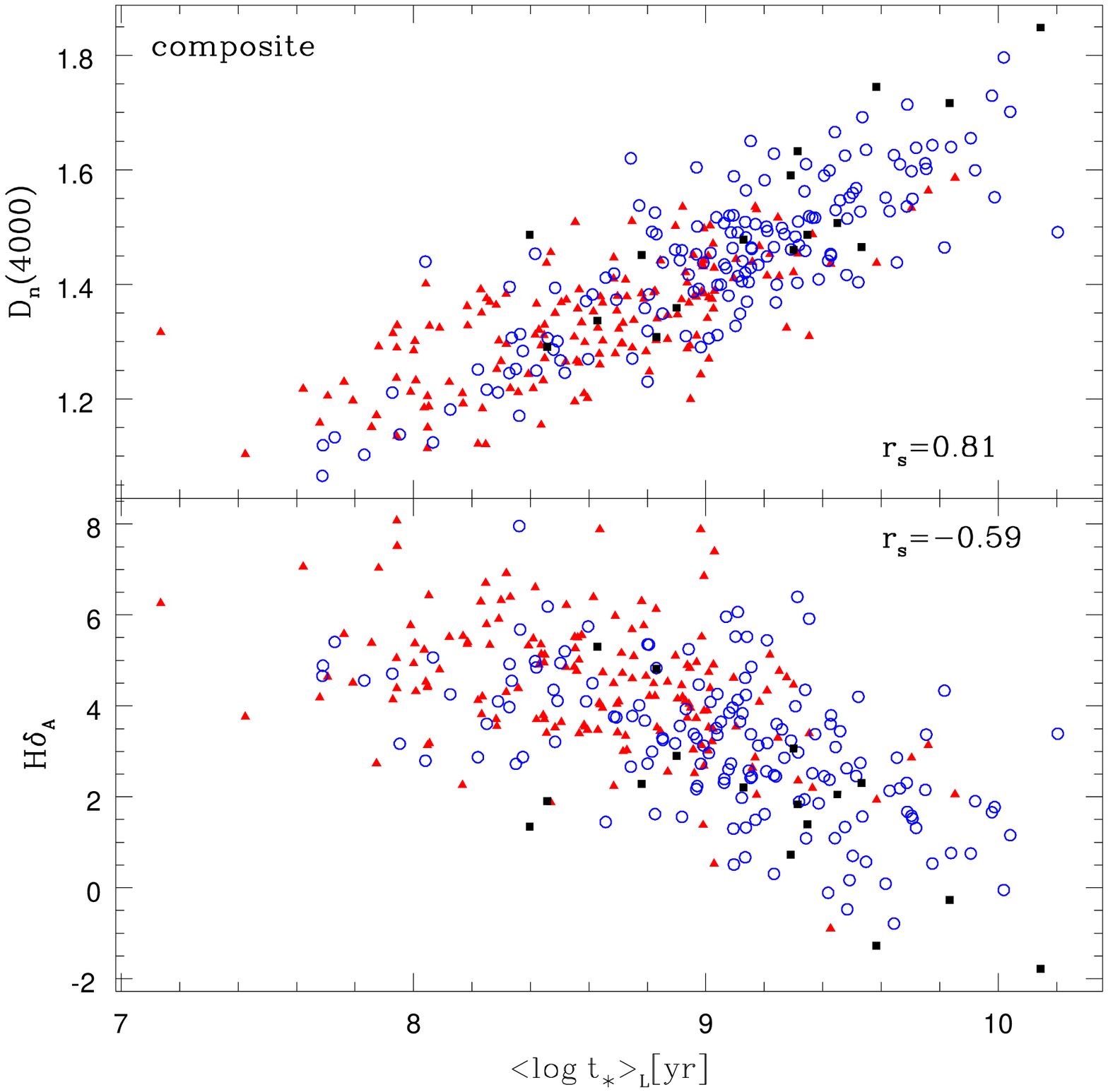}
   \includegraphics[height=6cm,width=6cm]{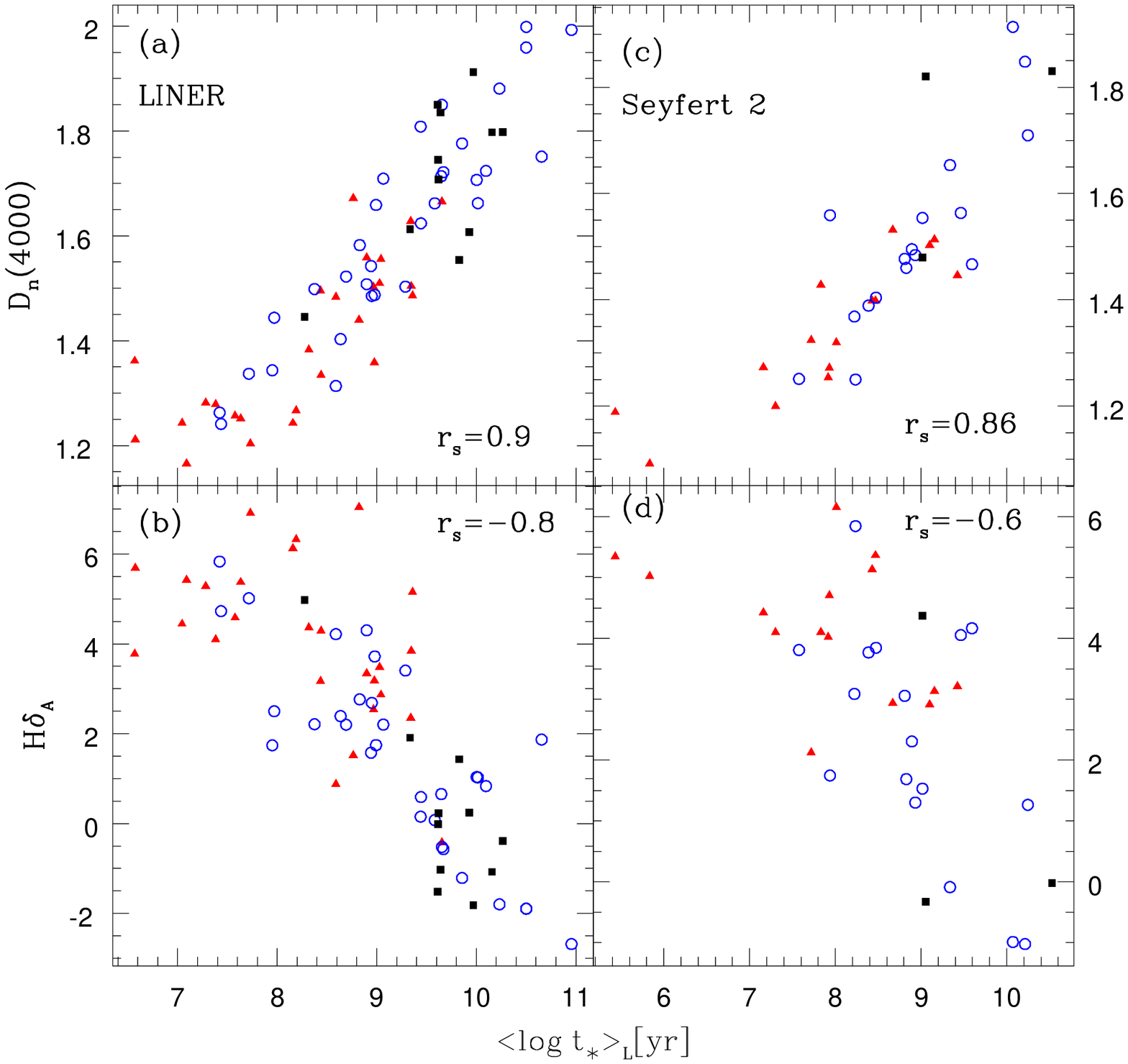}
   \caption{The relationships of light-weighted mean age
$\langle\,log\,t_{\ast}\rangle\,_{L}$ and $D_{n}(4000)$,
$H\delta_{A}$ for star-forming galaxies (left two panels),
composite galaxies (middle two panels), LINERs and Seyfert 2s (right four panels). The
Spearman-Rank Order correlation coefficient are shown in each
panel for all the sample galaxies there. All the symbols have the
same meanings as in Fig.~\ref{avneb_avstar.eps}.
           }
   \label{tl_d4000_hda.eps}
    \end{figure*}

%===============================================================
\subsection{$D_{n}(4000)$ and $H\delta_{A}$}
\label{sec.4.2}

We also acquired two age indicators from MPA/JHU database
directly: $D_{n}(4000)$ and $H\delta_{A}$.

$D_{n}(4000)$: The break at $4000 \AA$ is the strongest
discontinuity in the optical spectra and is available in our sample.
It was defined by Bruzual (1983) as the ratio of the average flux
density in the bands $4050-4250$ and $3750-3950$ $\AA$; however, we
used narrower continuum bands here (3850-3950 and 4000-4100 $\AA$),
which was introduced by Balogh et al. (1999). In this way the index
$D_{n}(4000)$ is less sensitive to reddening effects.  With the
increasing ages of the stellar populations of galaxies, their
$D_{n}(4000)$ values will increase as well, which means the higher
fraction of older populations.

$H\delta_{A}$: Strong $H\delta$ absorption lines arise in galaxies
that experienced a burst of star formation that ended $\sim$0.1-1
Gyr ago. The peak occurs once hot O and B stars, which have weak
intrinsic absorption, have terminated their evolution. The optical
light from the galaxies is then dominated by late-B to early-F
stars. Worthey \& Ottaviani (1997) defined an $H\delta_{A}$ index,
which contains a central bandpass ($4083-4122 \AA$ in MPA/JHU
database) bracketed by two pseudo-continuum band-passes.

The relationships between age indicators, $D_{n}(4000)$ and
$H\delta_{A}$, and the light-weighted mean age
$\langle\,log\,t_{\ast}\rangle\,_{L}$ for each subsample are
performed in Fig.~\ref{tl_d4000_hda.eps}. We also show the
Spearman-rank order correlation coefficient in each panel. In the
left two panels, we find there is a very strong correlation between
$D_{n}(4000)$ and our light-weighted mean age for star-forming
galaxies. Another age indicator, $H\delta_{A}$, presents a weaker
correlation (lower correlation coefficient) with mean stellar age
than does $D_{n}(4000)$. The weaker correlation between
$H\delta_{A}$ and the mean stellar age may be due to the fact that
$H\delta_{A}$ does not monotonously vary following the evolution of
the galaxy (Kauffmann et al. 2003b). Therefore, we suggest that
$D_{n}(4000)$ is a good age indicator as the scatter is small and
the relationship is tight in the left hand panel of
Fig.~\ref{tl_d4000_hda.eps}.

For composite galaxies (middle two panels in
Fig.~\ref{tl_d4000_hda.eps}), we draw a similar conclusion to that
of star-forming galaxies: $D_{n}(4000)$ is a good age indicator,
as the correlation coefficient is high and the diffusion is small.
Although $H\delta_{A}$ performs better than that of star-forming
galaxies, there are still some diffusions in the figure.

The right four panels show the results of LINERs and Seyfert 2s: the
left two for LINERs and the right two for Seyfert 2s. In the left
two  we find there are strong correlations between $D_{n}(4000)$,
$H\delta_{A}$, and $\langle\,log\,t_{\ast}\rangle_{L}$ for LINERs,
while from the right two we find $D_{n}(4000)$ is still a good age
indicator for Seyfert 2s. In view of so few data points contained in
LINERs and Seyfert 2s, especially Seyfert 2s, we cannot draw any
absolute conclusion.

Figure~\ref{tl_d4000_hda.eps} generally shows that the ULIGs and
LIGs have relatively lower light-weighted mean ages than others,
which is consistent with our results above (Sect. 3.3). In other
words, we have tested all individual galaxies, so that we can
confirm the robustness of the results from fitting the combined
spectra.
%-------------------------------------------------------------figure11
  \begin{figure*}
   \centering
   \includegraphics[height=6cm,width=6cm]{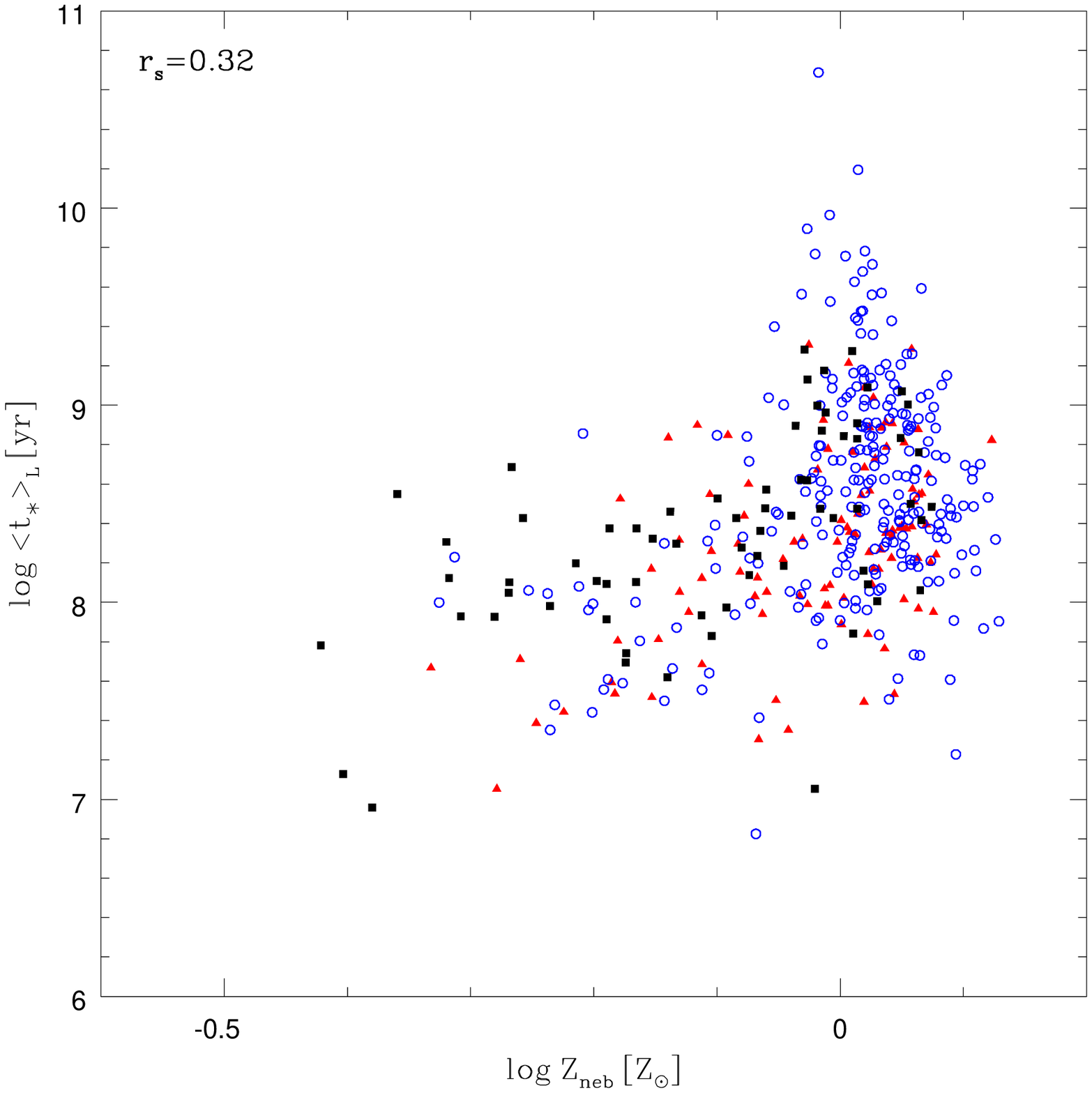}
   \includegraphics[height=6cm,width=6cm]{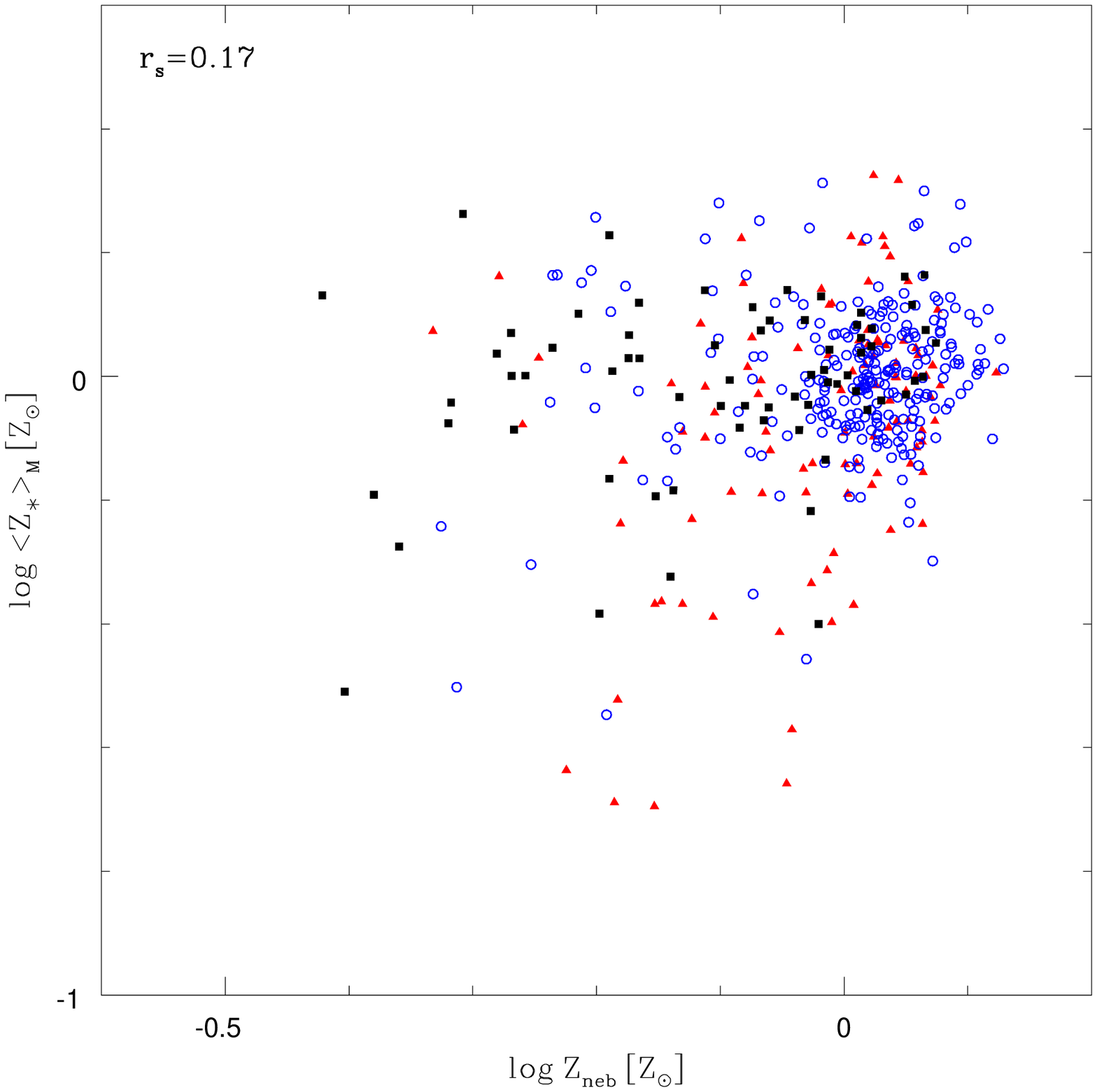}
   \caption{The relationships between the nebular metallicity $Z_{neb}$ and
the light-weighted mean age $\langle\,log\,t_{\ast}\rangle\,_{L}$
(left panel), between $Z_{neb}$ and the mass-weighted mean stellar metallicity
$\langle\,Z_{\ast}\rangle\,_{M}$ (right panel) for star-forming
galaxies. The Spearman-rank order correlation coefficient are shown
in each panel for all the sample galaxies there. All the symbols
have the same meanings as in Fig.~\ref{avneb_avstar.eps}.
            }
   \label{zneb_tl_zm.sf.sta06.eps}
    \end{figure*}
%=================================================================
\subsection{Nebular metallicity and mean stellar age, mean stellar metallicity}
\label{sect.4.3}

In this section, we show the distributions between nebular
metallicity $Z_{neb}$ and the light-weighted mean stellar age
$\langle\,log\,t_{\ast}\rangle\,_{L}$ (Eq.~\ref{tl}), the
mass-weighted mean stellar metallicity
$\langle\,Z_{\ast}\rangle\,_{M}$ (Eq.~\ref{zm}) provided by
STARLIGHT. We only show the results of star-forming galaxies for two
reasons, one of which is the uncertainty of the estimation of
$Z_{neb}$ for composite galaxies and AGNs. Another reason is that we
want to compare our results with that of Asari et al. (2007), whose
sample only includes all the star-forming galaxies.

After correcting all of the useful line fluxes  for extinction, we
quantified the nebular metallicity $Z_{neb}$ by the oxygen abundance
of the star-forming galaxies. To be consistent with Asari et al.
(2007), the nebular metallicity of star-forming galaxies was
obtained according to Eq.~(\ref{zo3n2.sf}) where
$O_{3}N_{2}$=log\,([OIII]$\lambda5007$/[NII]$\lambda$6583)
(Stasi\'{n}ska et al. 2006; Pettini et al. 2004; Liang et al. 2006):
\begin{equation}
\label{zo3n2.sf} \rm
log\,Z_{neb}^{o3n2}={log(O/H)}/{log(O/H)_{\odot}}=-0.14-0.25\times
O_{3}N_{2}\,[Z_{\odot}],
\end{equation}
where (O/H)$_{\odot}$=4.9$\times$10$^{-4}$.

 In Fig.~\ref{zneb_tl_zm.sf.sta06.eps} we show the
distributions between $Z_{neb}$ and
$\langle\,log\,t_{\ast}\rangle\,_{L}$ (the left hand panel),
$\langle\,Z_{\ast}\rangle\,_{M}$ (the right hand panel). From the
left hand panel, we find there is a correlation with intermediate
strength, and the correlation coefficient is 0.32. It seems that the
young stellar population makes a relatively dominant contribution in
low metallicity galaxies. The right hand panel shows that there is a
weak correlation between the mean stellar metallicity and the
nebular metallicity, and the correlation coefficient is only 0.17.
The chemical enrichment histories of star and gas are correlated on
the basis of simple chemical evolution scenarios, while the weakness
and scatter of this relation may be due to either the disturbance
from interactions/mergers and the indefinability of our calculations
or the intrinsically different evolutionary phases (Asari et al.
2007). These two relations are similar to those of the total
star-forming galaxies presented in the Fig.5 in Asari et al. (2007).
%===============================================
\section{Summary and conclusions}

Based on a large sample (849) of local infrared galaxies selected
from the cross-identification of the SDSS DR4 and IRAS PSCz, we fit
the spectral absorptions and continua of the sample galaxies to
study their stellar populations by using the SSPs of BC03 and
spectra of star clusters as bases and the software STARLIGHT. The
sample galaxies are divided into four or three subgroups by
following their emission-line ratios or infrared luminosity. First,
we investigated the effects from spectral class (star-forming,
composite galaxies, LINERs, and Seyfert 2s) and infrared luminosity
(ULIGs \& LIGs, starbursts, normal galaxies) on stellar populations.
Second, we fit our sample galaxies with SSPs from BC03 and spectra
of star clusters. Finally, we compared the stellar extinction, mean
stellar age, and mean stellar metallicity derived from spectral
synthesis with nebular extinction, nebular metallicity, and two age
indicators, $D_{n}(4000)$ and $H\delta_{A}$, obtained from the
MPA/JHU catalog.

As to the contributions to light, we found that LINERs presented the
oldest stellar populations among 4 spectral classes, and the
importance of young populations was decreasing from star-forming,
composite galaxies, Seyfert 2s, to LINERs. The dominant
populations of star-forming and composite galaxies are those with
metallicity $Z=0.2Z_{\odot}$, and the LINERs and Seyfert 2s are more
metal-rich. While ULIGs \& LIGs present the youngest populations
among the star-forming galaxies in 3 infrared luminosity bins. 
The normal galaxies are more metal-rich than the ULIGs \& LIGs and
starbursts. However, the dominant contributors to mass are all old
stellar populations, and the spectral class and the infrared
luminosity have no effect on this result. The results of fittings by
using spectra of star clusters as bases are consistent with those
from the SSPs of BC03.

The stellar and nebular extinctions are correlated in all sample
galaxies, and the ionized gas suffers twice as much extinction as
 a star. The $D_{n}(4000)$ is a much better age indicator than
$H\delta_{A}$, which does not monotonously vary following the
evolution of the galaxy. When we compared nebular metallicity with
other parameters for star-forming galaxies, we found that the mean
stellar age is correlated with nebular metallicity in an
intermediate strength, and $Z_{neb}$ is weakly correlated with
$\langle\,Z_{\ast}\rangle\,_{M}$.

%-----------------------------------------------------------------------
\begin{acknowledgements}
We thank our referee for the valuable
  comments and suggestions, which helped improve this work.
  We thank Philippe Prugniel for very helpful discussions,
  and thank James Wicker for kindly correcting the text.
We acknowledge the support of the Natural Science Foundation of
China (NSFC) Foundation under Nos.10403006,
 10433010, 10673002, 10573022, 10333060, and 10521001,
and the
National Basic Research Program of China (973 Program) No.2007CB815404,06.
This work is supported by the Knowledge Innovation Program of the Chinese Academy of Sciences.
The STARLIGHT project is supported by the Brazilian agencies CNPq,
 CAPES, and FAPESP and by the France-Brazil CAPES/Cofecub program.
 We thank the wonderful SDSS and IRAS database.
\end{acknowledgements}

\clearpage

\begin{thebibliography}{b}

  \bibitem[2006]{adelman} Adelman-McCarthy, J. K. et al. 2006, \apjs, 162, 38
  \bibitem[1993]{alongi} Alongi, M. et al. 1993, \aaps, 97, 851
  \bibitem[2007]{asari} Asari, N. V. et al. 2007, \mnras, 381, 263
  \bibitem[1981]{baldwin} Baldwin, J. A., Phillips, M. M. \& Terlevich, R.  1981, \pasp, 93, 5
 \bibitem[1999]{balogh} Balogh, M. L., Morris, S. L., Yee, H. K. C., Carlberg, R. G. \& Ellingson, E. 1999, \apj, 527, 54
  \bibitem[2003]{bernardi} Bernardi, M. et al. 2003,  \aj, 125, 1817
\bibitem[1986]{bica} Bica, E. \& Alloin, D. 1986a, \aap, 162, 21
\bibitem[1986]{bica} Bica, E. \& Alloin, D. 1986b, \aaps, 66, 171
\bibitem[2000]{boisson} Boisson, C., Joly, M., Moultaka, J., Pelat, D. \& Roos, M. S. 2000, \aap, 357, 850
\bibitem[2004]{boisson} Boisson, C., Joly, M., Pelat, D. \& Ward, M. J. 2004, \aap, 428, 373
\bibitem[2000]{bonatto} Bonatto, C., Bica, E., Pastoriza, M. G. \& Alloin, D. 2000, \aap, 355, 99
  \bibitem[1993]{bressan} Bressan, A., Fagotto, F., Bertelli, G. \& Chiosi, C. 1993, \aaps, 100, 647
  \bibitem[2004]{brinchmann} Brinchmann, J. et al. 2004,  \mnras, 351, 1151
 \bibitem[1983]{bruzual} Bruzual, A. G. 1983, \apj, 273, 105
  \bibitem[2003]{bruzual}Bruzual, A. G. \& Charlot, S. 2003, \mnras, 344, 1000
 \bibitem[1994]{calzetti} Calzetti, D., Kinney, A. L. \& Storchi-Bergmann, T. 1994, \apj, 429, 582
  \bibitem[2000]{calzetti} Calzetti, D. et al. 2000,  \apj, 533, 682
  \bibitem[2006]{cao} Cao C. et al. 2006, \cjaa, 6, 197
  \bibitem[1989]{cardelli}  Cardelli, J. A., Clayton, G. C. \& Mathis, J. S. 1989, \apj, 345, 245
  \bibitem[2003]{chabrier} Chabrier, G., 2003, \pasp, 115, 763
\bibitem[1996]{charlot} Charlot, S., Worthey, G. \& Bressan, A. 1996, \apj, 457, 625
\bibitem[2008]{chen} Chen, X. Y., Hao, C. N. \& Wang, J.2008, \cjaa, 8, 25
\bibitem[2001]{cid fernandes} Cid Fernandes, R., Sodr\'{e}, L., Schmitt, H. R. \& Le\~{a}o, J. R. S. 2001, \mnras, 325, 60
\bibitem[2003]{cid fernandes} Cid Fernandes, R., Le\~{a}o, J. R. S. \& Lacerda, R. R. 2003, \mnras, 340, 29
\bibitem[2004]{cid fernandes} Cid Fernandes, R. et al. 2004a, \apj, 605, 105
 \bibitem[2004]{cid fernandes} Cid Fernandes, R. et al. 2004b, \mnras, 355, 273
\bibitem[2005]{cid fernandes} Cid Fernandes, R., Delgado, R. M. G., Storchi-Bergmann, T., Martins, L. P. \& Schmitt, H. 2005a, \mnras, 356, 270
 \bibitem[2005]{cid fernandes} Cid Fernandes, R., Mateus, A. \& Sodr\'{e}, L., Stasi\'{n}ska, G. \& Gomes, J. M. 2005b, \mnras, 358, 363
\bibitem[2007]{cid fernandes} Cid Fernandes, R. 2007, in press [astro-ph/0701902]
\bibitem[2008]{cid fernandes} Cid Fernandes, R. et al. 2008, in press [astro-ph/0802.0849]
\bibitem[1999]{elbaz} Elbaz, D. et al. 1999, \aap, 351, L37
  \bibitem[2002]{elbaz} Elbaz, D. et al. 2002, \aap, 384, 848
\bibitem[1972]{faber} Faber, S. M. 1972, \aap, 20, 361
  \bibitem[1994]{fagotto} Fagotto, F., Bressan, A., Bertelli, G. \& Chiosi, C. 1994a, \aaps, 104, 365
  \bibitem[1994]{fagotto} Fagotto, F., Bressan, A., Bertelli, G. \& Chiosi, C. 1994b, \aaps, 105, 29
\bibitem[1999]{flores} Flores, H. et al. 1999, \apj, 517, 148
  \bibitem[1996]{fukugita} Fukugita, M. et al. 1996,  \aj, 111, 1748
  \bibitem[1996]{girardi} Girardi, L., Bressan, A., Chiosi, C., Bertelli, G. \& Nasi, E. 1996, \aaps, 117, 113
\bibitem[2001]{gonzalez} Gonz\'{a}lez Delgado, R. M., Heckman, T. \& Leitherer, C. 2001, \apj, 546, 845
\bibitem[2004]{gonzalez} Gonz\'{a}lez Delgado, R. M. et al. 2004, \apj, 605, 127
  \bibitem[2005]{greene}  Greene, J. E. \& Ho L. C. et al. 2005, \apj, 627, 721
  \bibitem[2006]{greene}  Greene, J. E. \& Ho L. C. et al. 2006, \apj, 641, 117
\bibitem[2001]{hammer} Hammer, F., Gruel, N., Thuan, T. X., Flores, H. \& Infante, L. 2001, \apj, 550, 570
\bibitem[2005]{hammer} Hammer, F. et al. 2005, \aap, 430, 115
  \bibitem[2004]{heckman} Heckman, T. M. et al. 2004, \apj, 613, 109
  \bibitem[1988]{helou} Helou, G., Khan, I. R., Malek, L. \& Boehmer, L. 1988,  \apjs, 68, 151
\bibitem[2001]{joguet} Joguet, B., Kunth, D., Melnick, J., Terlevich, R. \& Terlevich, E. 2001, \aap, 380, 19
  \bibitem[2003]{kauffmann} Kauffmann, G. et al. 2003a, \mnras, 346, 1055
\bibitem[2003]{kauffmann} Kauffmann, G. et al. 2003b, \mnras, 341, 33
  \bibitem[2001]{kewley} Kewley, L. J., Dopita, M. A., Sutherland, R. S., Heisler, C. A. \& Tervena, J. 2001, \apj, 556, 121
\bibitem[2006]{kewley} Kewley, L. J., Groves, B., Kauffmann, G. \& Heckman, T. 2006, \mnras, 372, 961
\bibitem[2008]{koleva} Koleva, M., Prugniel, Ph., Ocvirk, P., Le Borgne, D. \& Soubiran, C. 2008, \mnras, in press [astro-ph/0801.0871]
\bibitem[2003]{kong} Kong, X., Charlot, S., Weiss, A. \& Cheng, F. Z. 2003,  \aap, 403, 877
\bibitem[1978]{koski} Koski, A. T. 1978, \apj, 223, 56
 \bibitem[2003]{le borgne} Le Borgne, J.-F. et al. 2003, \aap, 402, 433
\bibitem[2005]{le floch} Le Floc'h, E. et al. 2005, \apj, 632, 169
\bibitem[2004]{liang} Liang, Y. C. et al. 2004, \aap, 423, 867
 \bibitem[2006]{liang} Liang, Y. C. et al. 2006, \apj, 652, 257
  \bibitem[2006]{mateus} Mateus, A. et al. 2006, \mnras, 370, 721
\bibitem[1989]{osterbrock} Osterbrock, D. E., 1989, Astrophysics of Gaseous
Nebulae  and Active Galactic Nuclei (Mill Valley: University Science Books)
\bibitem[2007]{papovich} Papovich, C. et al. 2007, \apj, 668, 45
\bibitem[2004]{pettini} Pettini, M. \& Pagel, B. E. J. 2004, \mnras, 348, L59
\bibitem[2000]{raimann} Raimann, D., Bica, E., Storchi-Bergmann, T., Melnick, J. \& Schmitt, H. 2000, \mnras, 314, 295
\bibitem[2008]{riffel} Riffel, R., Pastoriza, M. G., Rodr\'{i}guez-Ardila, A. \& Maraston, C. 2008, \mnras, in press [astro-ph/0805.1167]
\bibitem[2006]{sanchez} S\'{a}nchez-Bl\'{a}zquez, P. et al. 2006, \mnras, 371, 703
\bibitem[1996]{sanders} Sanders, D. B. \& Mirabel, I. F. 1996, \araa, 34, 749
  \bibitem[2000]{saunders} Saunders, W. et al. 2000, \mnras, 317, 55
\bibitem[1998]{schlegel} Schlegel, D. J., Finkbeiner, D. P. \& Davis, M. 1998, \apj, 500, 525
\bibitem[1996]{schmitt} Schmitt, H. R., Bica, E. \& Pastoriza, M. G. 1996, \mnras, 278, 965
\bibitem[1999]{schmitt} Schmitt, H. R., Storchi-Bergmann, T. \& Cid Fernandes, R. 1999, \mnras, 303, 173
\bibitem[1979]{seaton}Seaton, M. J., 1979, \mnras, 187, 73
\bibitem[2002]{smith} Smith, J. A. et al. 2002, \aj, 123, 2121
  \bibitem[1981]{shuder} Shuder, J. M. \& Osterbrock, D. E. 1981, \apj, 250, 55
\bibitem[1987]{soifer} Soifer, B. T., Houck, J. R. \& Neugebauer, G. 1987, \araa, 25, 187
 \bibitem[2004]{stasinska} Stasi\'{n}ska, G., Mateus, A., Sodr\'{e}, L. \& Szczerba, R. 2004, \aap, 420, 475
 \bibitem[2006]{stasinska} Stasi\'{n}ska, G., 2006, \aap, 454, L127
\bibitem[2008]{stasinska} Stasi\'{n}ska, G. et al. 2008, \mnras, in press [astro-ph/0809.1341]
\bibitem[2000]{storchi} Storchi-Bergmann, T., Raimann, D., Bica, E. L. D. \& Fraquelli, H. A. 2000, \apj, 544, 747
  \bibitem[2002]{stoughton} Stoughton, C. et al. 2002, \aj, 123, 485
  \bibitem[2002]{strauss} Strauss, M. A. et al. 2002, \aj, 124, 1810
\bibitem[1978]{tinsley} Tinsley, B. M. 1978, \apj, 222, 14
\bibitem[2004]{tremonti} Tremonti, C. A. et al. 2004, \apj, 613, 898
\bibitem[1999]{vazdekis} Vazdekis, A. 1999, \apj, 513, 224
\bibitem[1987]{veilleux}  Veilleux, S. \& Osterbrock, D. E. 1987, \apjs, 63, 295
  \bibitem[1995]{veilleux}   Veilleux, S., Kim, D.-C., Sanders, D. B., Mazzarella, J. M. \& Soifer, B.
  T., 1995, \apjs, 98, 171
\bibitem[2006]{wang} Wang, J. L. et al. 2006, \apj, 649, 722
\bibitem[2008]{wang} Wang, J. L. 2008, \cjaa, 8, 643
\bibitem[2004]{westera} Westera, P., Cuisinier, F., Telles, E. \& Kehrig, C. 2004, \aap, 423, 133
 \bibitem[1997]{worthey} Worthey, G. \& Ottaviani, D. L. 1997, \apjs, 111, 377
\bibitem[2004]{zheng} Zheng, X. Z., Hammer, F., Flores, H., Ass\'{e}mat, F. \& Pelat, D. 2004, \aap, 421, 847
\bibitem[2007]{zheng} Zheng, X. Z. et al. 2007, \apj, 670, 301
\end{thebibliography}
\end{document}